%% file: main-arxiv.tex
\documentclass[sigconf,balance=false]{acmart}

\acmYear{}
\acmVolume{}
\acmNumber{}
\acmDOI{XXXXXXX.XXXXXXX}
\acmISBN{}
\acmConference{}{}{}
\setcopyright{none}
\renewcommand\footnotetextcopyrightpermission[1]{} %

\usepackage{enumitem}

\usepackage{xspace}
\usepackage{algorithm}
\usepackage{algpseudocode}
\usepackage[most]{tcolorbox}
\tcbuselibrary{breakable}
\newcommand{\name}{\textsc{Narriva}\xspace}
\usepackage[textsize=footnotesize,color=green!40,
]{todonotes}
\usepackage[font=small,labelfont=bf]{caption}
\usepackage{subcaption}
\usepackage{multirow}
\usepackage{booktabs}
\usepackage{tabularx}

\newtcolorbox[auto counter, number within=section]{promptbox}[2][]{%
    breakable,                %
    sharp corners,            %
    colback=white,            %
    colframe=black,           %
    fontupper=\footnotesize,   %
    title={Prompt \thetcbcounter: #2}, %
    colbacktitle=white,       %
    coltitle=black,           %
    fonttitle=\bfseries\small, %
    attach title to upper,    %
    after title={\medskip\hrule\medskip}, %
    label={#1},               %
    before skip=10pt,         %
    after skip=10pt           %
}

\begin{document}

\title{Text-Based Personas for Simulating User Privacy Decisions}
\pagestyle{plain}

\author{Kassem Fawaz}
\authornote{Work done at Google as a visiting faculty researcher.}
\affiliation{%
  \institution{University of Wisconsin--Madison}
  \country{USA}}
\email{kfawaz@wisc.edu}

\author{Ren Yi}
\affiliation{%
  \institution{Google Research}
  \country{USA}}
\email{ryi@google.com}

\author{Octavian Suciu}
\affiliation{%
  \institution{Google Research}
  \country{USA}}
\email{osuciu@google.com}

\author{Rishabh Khandelwal}
\affiliation{%
  \institution{Google}
  \country{USA}}
\email{krishab@google.com}

\author{Hamza Harkous}
\affiliation{%
  \institution{Google}
  \country{Switzerland}}
\email{harkous@google.com}

\author{Nina Taft}
\affiliation{%
  \institution{Google}
  \country{USA}}
\email{ninataft@google.com}

\author{Marco Gruteser}
\affiliation{%
  \institution{Google Research}
  \country{USA}}
\email{gruteser@google.com}

\renewcommand{\shortauthors}{Fawaz et al.}

\begin{abstract}
\input{content/0_abstract}

\end{abstract}


\maketitle

\input{content/10_introduction}

\input{content/20_related}

\input{content/30_overview}

\input{content/40_system}

\input{content/50_eval}

\input{content/51_qual}

\input{content/60_discussion}

\input{content/70_conclusion}

\input{content/71_ethics}

\bibliographystyle{ACM-Reference-Format}
\bibliography{sample-base}

\clearpage
\appendix
\input{content/81_appendix}

\end{document}

%% file: content/0_abstract.tex
The ability to simulate human privacy decisions has significant implications for aligning autonomous agents with individual intent and conducting cost-effective, large-scale privacy-centric user studies. Prior approaches prompt Large Language Models (LLMs) with natural language user statements, data-sharing histories, or demographic attributes to simulate privacy decisions. These approaches, however, fail to balance individual-level accuracy, human auditability, token efficiency, and population-level representation. We present \name, an approach that generates text-based synthetic privacy personas to address these shortcomings. \name grounds persona generation in prior user privacy decisions, such as those from large-scale survey datasets, rather than purely relying on demographic stereotypes. It compresses this data into concise, human-readable summaries structured by established privacy theories. Through benchmarking across five diverse datasets, we analyze the characteristics of \name's synthetic personas in modeling both individual and population-level privacy preferences. We find that grounding personas in past privacy behaviors achieves up to 87\% predictive accuracy, improving over a non-personalized LLM baseline by 6-17 percentage points across datasets, while yielding an 80-95\% reduction in prompt tokens compared to in-context learning with raw examples. Finally, we demonstrate that personas synthesized from a single survey can reproduce the aggregate privacy behaviors and statistical distributions of entirely different studies.

%% file: content/10_introduction.tex
\section{Introduction}

Measuring and modeling users' privacy-related reactions, preferences, and behaviors is a long-standing challenge~\cite{acquisti2015privacy}. Addressing this challenge allows system developers, researchers, and other stakeholders to understand at scale how users make privacy-related decisions, perceive privacy risks in new technologies, react to privacy incidents, and utilize privacy-enhancing technologies. Recent developments in agentic frameworks make this challenge even more pressing. Autonomous agents will need to make data-sharing decisions on behalf of users in many contexts. While aligning autonomous actions with human intent is a broad challenge, it is important to measure how an agent's automated choices reflect a user's  privacy preferences~\cite{flemings2025}---a task central to the development of effective personalized privacy assistants.

The conventional approach to capturing and evaluating privacy decision-making relies on directly measuring user behaviors and attitudes at both the individual and population levels. These measurements take place through either controlled user studies (e.g., surveys, interviews, and lab experiments) or observational methods (e.g., on-boarding questions, recording app permissions, telemetry, or in-the-wild interactions). At the individual level, these methods capture the granular privacy preferences, concerns, and utility trade-offs, which can guide and evaluate personalized, automated decisions. At the population level, they allow researchers to identify privacy norms and measure how populations perceive and reason about privacy. 

While directly measuring user decision-making remains the gold standard for grounding privacy models, relying on it continuously is unscalable and costly~\cite{li2026llmagentssimulateenduser, anthis2025position}. At the population level, repeatedly testing new experimental conditions or design combinations to understand group norms requires large participant pools. This leads to prohibitively expensive studies, participant fatigue~\cite{groschupp2025}, and non-representative sample sizes~\cite{krsteski2025, suh2025}. At the individual level, initial surveys provide an essential baseline, but relying exclusively on direct measurement means users must be prompted every time a new context emerges. Therefore, using infrequent measurement to ground simulated decisions can reduce the friction and fatigue of constantly requiring updated input~\cite{groschupp2025}.

In light of these practical constraints, there is a rising interest in leveraging LLMs to simulate human decisions~\cite{suh2025,krsteski2025,anthis2025position} in social sciences~\cite{park2024simulations}, human-computer interaction (HCI)~\cite{LuUxAgent}, and privacy~\cite{groschupp2025,flemings2025,sullivan2025, li2026llmagentssimulateenduser}. Researchers have demonstrated and benchmarked LLMs' ability to predict privacy decisions at the individual level~\cite{groschupp2025,flemings2025,sullivan2025} and simulate population-level privacy phenomena~\cite{li2026llmagentssimulateenduser}. The key underlying approach is to use persona-based prompts to steer the model to answer questions from the perspective of a specific user profile. There are two methods for generating these personas. Individual-level personas are constructed by compiling granular data from a specific user. In the privacy domain, models are prompted with natural language user statements~\cite{groschupp2025} or histories of prior data-sharing decisions~\cite{flemings2025, sullivan2025}. On the other hand, population-level personas are synthesized to represent demographic segments or broad archetypes rather than a specific individual. These are typically constructed using demographic attributes (e.g., age, gender, income)~\cite{li2026llmagentssimulateenduser}.

However, existing approaches struggle to balance four competing dimensions: \textit{predictiveness} (accuracy in simulating individual decisions on novel questions), \textit{human auditability} (inherent semantic transparency of representing privacy logic in natural language rather than opaque logs), \textit{scalability} (maintaining bounded context lengths), and \textit{representation} (authentically capturing population diversity without relying on generic LLM stereotypes). For instance, constructing individual-level models by concatenating raw histories into an LLM's context window~\cite{flemings2025} yields high predictiveness when novel queries match the format of the historical data. However, this approach struggles to generalize to new domains or question types, and maintaining a growing raw history affects scalability and auditability. Conversely, relying on user-provided natural language statements (e.g., ``I am very careful about sharing information'') offers high auditability but lower predictiveness, especially in novel scenarios~\cite{groschupp2025}. Finally, synthesizing pre-generated demographic personas~\cite{Farquhar2024} is highly scalable but, as highlighted by Anthis et al.~\cite{anthis2025position} and Li et al.~\cite{li2026llmagentssimulateenduser}, often falls short in representation and predictiveness, producing biased outputs that rely on LLM-specific artifacts rather than authentic human nuances~\cite{krsteski2025}. 

We systematically investigate whether text-based personas can address the challenge of balancing predictiveness, auditability, scalability, and representation, by answering the following research questions:

\begin{itemize}[leftmargin=*]

    \item \textbf{RQ1.}  Can text-based personas effectively model individual-level preferences when simulating personalized privacy-related decisions?
    
    \item \textbf{RQ2.}  Can concise, text-based synthetic personas capture real-world population characteristics to accurately reproduce population-level privacy perspectives?

    \item \textbf{RQ3.}  Are these generated personas sufficiently general to predict population-level perspectives in completely independent studies?

    \end{itemize}

To tackle these research questions, we propose \name (derived from \textbf{Nar}rative Personas for P\textbf{riva}cy), a framework designed to successfully balance these four dimensions. 
To achieve \textit{auditability and scalability}, \name compresses raw privacy data into generalized, text-based privacy personas, shifting from uninterpretable raw data logs to concise, human-readable narratives. To ensure these personas maintain population-level \textit{representation}, we ground their generation in large-scale real-world survey datasets rather than relying on LLM-generated demographic stereotypes. Finally, to achieve individual-level \textit{predictiveness}, \name employs an iterative optimization process that tests the generated persona against the user's historical decisions, refining the narrative until it can consistently simulate that individual's privacy preferences.

We apply \name to five privacy-centric datasets---from the Pew Research Center (PP1, W49, W127), ACM CHI 2021 (CAuthN), and PETS Symposium 2024 (SPA)---to evaluate the performance of the text-based personas. Our main findings are summarized below:

\begin{itemize}[leftmargin=*]
\item \textbf{[RQ1 \& RQ2] Text-based personas match the accuracy of raw input while being more concise.} \name successfully condenses long question-answer histories into narratives up to 20x smaller. This reduction lowers inference compute costs, enhances human readability, and improves privacy by abstracting away exact historical interactions without degrading individual or population-level predictability.

\item \textbf{[RQ1] Grounding and Structure can enhance predictiveness of personas at the individual level.} We observe that grounding personas in behavioral questions makes them more predictive of practical decisions. Further, we find that representing personas through different privacy frameworks (e.g., Privacy Calculus, Bounded Rationality) offers more control over persona structure and can improve personalized predictions. With these enhancements, text-based personas can have up to 87\% accuracy in predicting data-sharing decisions in the SPA dataset.

\item \textbf{[RQ3] High-fidelity personas generalize across surveys but degrade over long temporal gaps.} Personas synthesized from one study can accurately simulate population-level responses for an independent study conducted a few years apart. Yet, transferability degrades for surveys separated by nearly a decade, reflecting shifts in the technological landscape and population attitudes.
\end{itemize}

We supplement our evaluation with a qualitative analysis of highly predictive and non-predictive personas. This analysis, coupled with our quantitative results, yields several insights to inform future research on simulating privacy decision-making. We find that several factors influence persona quality, including the nature of the underlying questions, the consistency of the user's decisions, their views toward institutions, and their perceived level of control. Future frameworks should account for these factors by grounding personas in precise, context-specific questions. Furthermore, these frameworks must incorporate mechanisms to detect when a user's preferences are unclear. In such moments of uncertainty, agents can proactively engage the user by highlighting personal inconsistencies and asking targeted questions to further refine the persona.

%% file: content/20_related.tex
\section{Related Work}
\label{sec:related}

Owing to their advanced reasoning capabilities, recent research has explored using LLMs to simulate human behaviors and decisions in several domains, including privacy. In a recent position paper, Anthis et al. argue that LLMs can simulate a spectrum of user studies, provided five key challenges are addressed~\cite{anthis2025position}. These challenges include diversity, bias, sycophancy, alienness, and generalization. Related works fall into two main categories: those that attempt to simulate general populations through synthetic personas, and those that simulate particular individuals from existing information about them.

\subsection{Population-Level Simulations}

The simulation of users in general user experience (UX) research settings is gaining popularity. Lu et al. propose an approach to pre-generate a set of user personas~\cite{LuUxAgent}. These personas include demographic attributes, such as age, gender, and income. Then, a browser-use agent is prompted with the user persona to perform a set of actions on a web interface to measure its usability. This process provides qualitative feedback for researchers to review. More recently, Bui et al. address the problem of how to guide LLMs to mimic the responses of a target population~\cite{bui-etal-2025-mixture}. To improve simulation diversity, they propose an agentic framework that dynamically samples a probabilistic mixture of personas, achieving better population alignment than existing baselines.

While synthetic personas show promise for general UX tasks, adapting them to the privacy domain remains challenging. Notably, concurrent work by Li et al. highlights the potential and the limitations of using out-of-the-box LLMs to simulate privacy and security behaviors~\cite{li2026llmagentssimulateenduser}. They introduce SP-ABCBench to evaluate the population-level alignment of LLMs using synthetically generated, demographic-based personas. They found that while population-level prompting works for broad user interactions, current models struggle with the nuance required for privacy, achieving moderate fidelity (scoring 50--64\% on average).

\subsection{Individual-level Simulation}

In the second category, researchers propose approaches to simulate individuals by grounding models in their prior data. In the broader social sciences, Park et al. perform in-depth interviews with more than 1000 individuals and propose an agentic framework to simulate their behavior~\cite{park2024simulations}. Their framework takes an entire, adaptively elicited interview transcript as context to answer questions about the user, reproducing answers to the General Social Survey with 85\% accuracy. Other works seek to align individual LLM simulations through fine-tuning. Suh et al. develop a fine-tuning method to align the probability distributions of responses from an LLM to those from human subpopulations~\cite{suh2025}. Krsteski et al. demonstrate that synthesized personas and basic fine-tuning often yield biased responses, proposing a post-hoc rectification mechanism using a small portion of user surveys to estimate debiasing parameters~\cite{krsteski2025}. Similarly, Wang et al. propose an optimal allocation rule to distribute limited human data between fine-tuning and rectification~\cite{wang2025efficient}.

In the privacy domain, several works explore the use of LLMs to guide personalized privacy decisions. Groschupp et al.~\cite{groschupp2025} investigate the capacity of LLMs to automate personalized privacy decisions for app permissions. By incorporating natural language user statements, such as ``I am very careful about sharing information,'' into the system prompt, their model achieves up to 88\% alignment with user decisions across the population. Similarly, Flemings et al.~\cite{flemings2025} employ logic-based inference on Contextual Integrity (CI) information flows to determine if past sharing decisions can predict future ones. In robotics, Sullivan et al.~\cite{sullivan2025} demonstrate that including a history of user decisions in the LLM prompt significantly improves alignment with future household robot interactions.

\subsection{Shortcomings}

Existing approaches to generating simulated privacy personas struggle to simultaneously achieve predictiveness, auditability, scalability, and representation. Population-level approaches rely on zero-shot demographic prompting. While they are highly scalable, they cannot represent the nuances of privacy decision-making of the underlying populations. They are also inherently incapable of predicting individual-level decisions. Individual-level methods that rely on raw, extensive decision logs or entire transcripts offer higher predictiveness for specific scenarios but inherently limit scalability and auditability. Finally, methods relying on brief, free-form natural language statements offer high auditability but do not generalize to novel scenarios. \name addresses these shortcomings by generating concise text-based personas that are predictive and grounded in actual user histories.

%% file: content/30_overview.tex
\section{Problem Statement}
\label{sec:problem}

\name generates synthetic privacy personas that help simulate human-like privacy decision-making.

\subsection{Objectives and Approaches}
The overall objective of \name is to generate a large set of synthetic text-based personas to enable the rigorous simulation and prediction of privacy decisions. It aims to balance the four dimensions of text-based personas--predictiveness, auditability, representation, and scalability as follows.

\paragraph{Predictiveness}
\name ensures that personas are \textit{predictive} of individual decisions by grounding them in the known history of the user, such as their prior privacy choices or answers to previous questionnaires. Rather than relying on a single summarization pass, this grounding happens through an iterative and lightweight optimization process. This process refines the generated persona until it can consistently predict the user's responses to privacy-related questions. 

\paragraph{Auditability and Scalability}
\name compresses raw data about an individual into concise, structured personas that provide \textit{semantic transparency and editability}, properties essential for establishing user trust in future personalized privacy assistants. In particular, it grounds these personas in established privacy decision-making frameworks (e.g., Privacy Calculus or Bounded Rationality), thereby standardizing their representation and ensuring that the underlying privacy logic is explicitly articulated. Rather than relying on unbounded interaction logs, \name explicitly extracts specific information from a user's history to populate a short, theory-driven template that remains scalable for subsequent inferences.

\paragraph{Representation}
\name ensures that the generated personas are \textit{representative} of a target population by applying its predictive and theory-driven persona generation to privacy datasets. Privacy researchers have consistently measured user concerns, knowledge, attitudes, preferences, and behaviors related to privacy. Typically, this data is collected using careful procedures to sample individuals that represent a target population, especially in terms of demographic attributes. For example, the Pew Research Center samples US individuals, employing a sampling strategy to proportionately cover all subgroups in the US population. Synthesizing personas based on the historical responses of specific individuals from these datasets results in a set of personas that capture population-level decision-making.

\subsection{Problem Formalization}

We are given a dataset $S$ containing privacy-related responses from a set of individuals. This setting generalizes the case of simulating the decisions for a single individual. We represent this dataset as an $M \times N$ matrix of $M$ individuals and $N$ questions, such that the set of individual responses (rows) is $\mathcal{R} = \{\mathbf{r}_1, \mathbf{r}_2, \dots, \mathbf{r}_M\}$ and the set of questions (columns) $\mathcal{Q} = \{q_1, q_2, \dots, q_N\}$. Each question has a set of possible answers, which depend on its nature (e.g., binary, Likert, categorical). Let $A_i = \{a_{i,1}, a_{i,2}, \dots, a_{i,m_i}\},$ be the set of  $m_i$ possible answers for question $q_i$. For simplicity, we do not consider empty answers or refusals as possible answers. We also assume that there is a function $f_i$ for a question $q_i$ that consistently maps each answer to a numerical value: $f_i: A_i \rightarrow \{1, 2, \dots, m_i\}$ (e.g., $\{0,1\}$ for a binary question).

The response set $\mathbf{r}_j$ is defined as the vector of answers provided by the $j$-th individual, denoted as $\mathbf{r}_j = (r_{j,1}, r_{j,2}, \dots, r_{j,N})$, where each element $r_{j,i} \in A_i$. We further assume that the set of questions $\mathcal{Q}$ can be partitioned into a set of $K$ mutually exclusive domains $\mathcal{D} = \{D_1, D_2, \dots, D_K\}$, such that $\bigcup_{k=1}^K D_k = \mathcal{Q}$. Examples of domains include demographics, privacy decisions/behaviors, and privacy attitudes.

We additionally partition the set of questions $\mathcal{Q}$ into two disjoint subsets: a generation set $\mathcal{Q}_{gen}$ and an evaluation set $\mathcal{Q}_{eval}$. These subsets satisfy $\mathcal{Q}_{gen} \cup \mathcal{Q}_{eval} = \mathcal{Q}$ and $\mathcal{Q}_{gen} \cap \mathcal{Q}_{eval} = \emptyset$. For each response set $\mathbf{r}_j$, we define two subsets of responses. The first subset of responses, $\mathbf{r}_j^{\text{gen}}$, is used for generation and consists of answers to questions in $\mathcal{Q}_{gen}$. The second is used for evaluation:  $\mathbf{r}_j^{\text{eval}}$, which consists of answers to questions in $\mathcal{Q}_{eval}$.

\name has two methods. The generation method, $$ p_j = \texttt{generate}(\mathbf{r}_j^{\text{gen}}),$$ maps a response set $\mathbf{r}_j$ to a persona $p_j$. The prediction method produces an answer to a question ($q$ with correct response $r$), given a persona ($p_j$), where: $$ \hat{r} = \texttt{predict}(p_j, q). $$

Now, we can define evaluation tasks that assess the performance of the generated personas at the individual level and at the population level. The individual-level evaluation task compares the predicted answers from the generated persona against the evaluation set $\mathbf{r}_j^{\text{eval}}$.
In particular, for a response set belonging to an individual $j$, the individual-level accuracy is:
$$\text{acc}_j = \frac{1}{|\mathcal{Q}_{eval}|}\sum_{q_i \in \mathcal{Q}_{eval}} {\mathbb{I} (\hat{r}_{j,i} = r_{j,i})},$$ where $\mathbb{I}$ is the indicator function and $\hat{r}_{j,i} = predict(p_j,q_i)$. The macro average accuracy over the entire dataset is simply: $$\text{acc}_{S} = \frac{1}{M} \sum_{j=1}^{M} \text{acc}_j.$$

The population-level evaluation task measures the distributional similarity between the predicted answers and real answers to a given question. To that end, we utilize the Total Variation Distance (TVD). For a given question $q_i$ and response $a$, we obtain the probability mass function of true responses as $P_i(a) = \frac{1}{M} \sum_{j=1}^{M} \mathbb{I}(f_i(r_{j,i}) = a)$ and that of predicted responses as $\hat{P}_i(a) = \frac{1}{M} \sum_{j=1}^{M} \mathbb{I}(f_i(\hat{r}_{j,i}) = a)$. The Total Variation Distance $TVD_i$ for question $q_i$ is then:$$TVD_i = \frac{1}{2} \sum_{a=1}^{m_i} | P_i(a) - \hat{P}_i(a) |.$$ For consistent presentation between the individual and population level metrics, we also report the TVComplement from the Synthetic Data Vault~\cite{sdmetrics_tvcomplement}, which is the complement of the Total Variation Distance (TVD) (i.e., $1 - \text{TVD}$). This metric quantifies the similarity between real and synthetic distributions~\cite{tvcompl_1, tvcompl_2}.

We also report the macro average of the Total Variation Distance over all evaluation questions as:$$\text{TVD}_{S} = \frac{1}{|\mathcal{Q}_{eval}|} \sum_{q_i \in \mathcal{Q}_{eval}} \text{TVD}_i.$$ The average TVComplement is given as $1 - \text{TVD}_{S}$.

%% file: content/40_system.tex
\section{System Design}

We present the design of \name, a system that generates text-based personas based on answers to prior questions and uses them to predict answers to unseen questions about privacy. We start with a high-level overview of \name, followed by detailed descriptions of its components.

\begin{figure}[t]
  \centering
  \includegraphics[width=\linewidth]{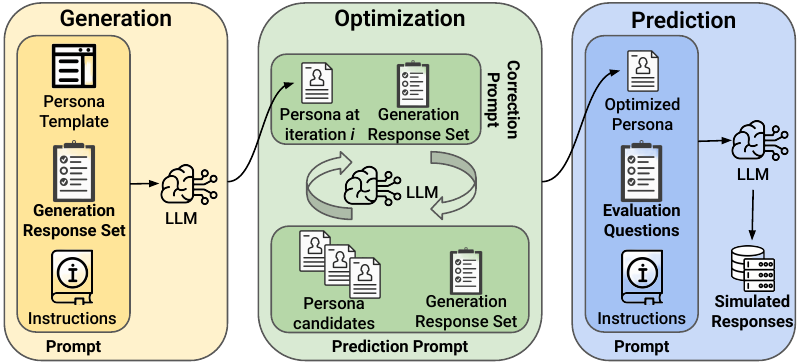}
  \small
  \caption{High-level overview of \name, which comprises three stages: initial persona generation, light-weight optimization, and prediction using the optimized persona.}

  \label{fig:overview}
\end{figure}

\subsection{High-level Overview}
At a high level (Fig.~\ref{fig:overview}), \name simulates user decisions on unseen privacy questions given their observed historical responses. It uses large language models (LLMs) to instantiate the $\texttt{generate}$ and $\texttt{predict}$ methods defined in Section~\ref{sec:problem}. 

\begin{itemize}[leftmargin=*]
    \item \textit{Generation:} \name prompts an LLM with a subset of an individual's responses ($\textbf{r}^\text{gen}$), a persona template, and instructions to generate the persona. The persona template offers a structured representation of the persona to be filled based on the individual's history.
    
    \item \textit{Prediction:} \name utilizes a prediction template to prompt an LLM to generate an answer to a given question. This template includes the generated persona, the question, the answer range, and the relevant instructions to simulate the user's decision.
\end{itemize}

To improve the accuracy of its simulations and practical scalability, \name employs an individual-level narrative optimization process. A user's privacy persona is highly specific and dynamically evolves as new questions are answered. To accommodate this dynamism, \name's individual-level optimization uses a lightweight and iterative loop to quickly refine the user's generated narrative for predictiveness and conciseness. This refinement anchors the personas in formal privacy theories, distilling a user's prior answers into structured dimensions like risk perception and perceived benefit rather than unstructured narratives. Once a persona has been generated, \name applies a pre-vetted prediction template to answer an unseen question. This design ensures \name remains agile when capturing individual behavioral shifts by optimizing the text persona itself, without requiring computationally prohibitive searches to continually rewrite the underlying prompt structure for every single user.

\paragraph{Why Optimize Text Personas?}
A naive approach to simulating privacy decisions inserts a user's entire historical response set ($\textbf{r}^\text{gen}$) directly into the generation model's prompt. While effective for static, one-off research experiments, this \textit{``raw-examples''} baseline fails to meet the scalability and auditability requirements of real-world privacy assistants.
First, as assistants continuously accumulate interactions, the raw context keeps growing. This growing context affects real-time latency and increases inference costs. While \name's narrative optimization requires periodic compute, this process can be run offline in the background with optimizations like caching and batching, keeping real-time inference costs low and bounded.
Second, maintaining a massive raw history of past actions makes it challenging for users to audit or edit their inferred preferences. It is unrealistic to expect users to manually sift through and edit raw historical logs to guide the LLM to act on their behalf.
Finally, raw logs  are tied to specific platform schemas and might contain unnecessary, granular details that inadvertently expose sensitive daily habits.

To address these issues, \name compresses a user's history into a readable text-based persona. This summarization approach keeps inference costs manageable over time. It naturally acts as a data minimization layer by abstracting away granular and privacy-sensitive decisions from raw logs. Furthermore, instead of relying on unstructured pattern-matching, it grounds these personas in established privacy frameworks. By standardizing the persona generation across users, these theoretical frameworks provide a predictable structure that is easier for users to audit and edit. In the following sections, we detail \name's three main components: initial persona generation, persona optimization, and prediction.

\subsection{Initial Persona Generation}
\label{sec:templates}
The first step in \name is generating an initial persona for an individual given their responses $\textbf{r}^\text{gen}$. As mentioned earlier, we employ templates to represent these personas in a more standard way. We derive these templates from earlier works that model privacy decision-making of individuals in different settings. 

These predefined templates serve two purposes in \name's design and evaluation. First, \name uses these templates to address the user experience aspects of text-based personas. Prompting the LLM to generate personas yields a free-form narrative that, while legible, can make it difficult for users to identify and edit specific privacy postures. Templates address this issue by organizing the persona into predefined and distinct sections that are clearly labeled, allowing users to quickly pinpoint and modify specific privacy traits. Ideally, \name should be able to utilize the structured persona template, without incurring a performance loss as compared to less structured baselines. Second, these templates allow us to assess the utility of structured, framework-driven personas against standard baseline and raw example-based prompting. We can thus empirically answer an important research question about whether representing a user through the lens of an established privacy framework enhances the model's ability to predict future privacy decisions.  We consider the following privacy frameworks, which are detailed in Appendix~\ref{appendix:privacy_theories}.

\begin{itemize}[leftmargin=*]
    \item \textbf{Basic Persona:} A baseline representation that uses chain-of-thought reasoning to analyze a user's responses and extract a concise narrative explaining their underlying logic.
    \item \textbf{Privacy Calculus:} Models individuals as rational actors by identifying their contextual variables, risk assessment methods, compelling benefit drivers, and strategies for weighing risks against benefits~\cite{privacy_calculus, calculus_1977, knijnenburg2017death}.
    \item \textbf{Bounded Rationality:} Accounts for cognitive limitations by identifying cognitive biases (e.g., optimism or present bias), emotional triggers, and primary heuristics that override a purely rational risk assessment~\cite{bounded_rational_1, bounded_rational_2}.
    \item \textbf{Protection Motivation Theory (PMT):} Frames privacy decisions as threat responses by evaluating an individual's risk appraisal (severity and vulnerability) alongside their coping appraisal (self-efficacy and effectiveness)~\cite{Rogers01091975}.
    
\end{itemize}

\subsection{Persona Optimization}
\input{content/41_opt_alog}

The core of \name is the persona optimization procedure, defined in Alg.~\ref{alg:persona_opt}. An initially generated persona, while conforming to a template, might not be completely predictive of the user's privacy decisions, attitudes, or preferences. As such, we develop an iterative optimization procedure to incrementally improve the generated persona, while imposing three constraints: predictiveness, conciseness, and generalizability. The optimization procedure takes as input a specific persona template, a set of generation questions ($\mathbf{q}^{\text{gen}}$) with their corresponding user responses ($\mathbf{r}^{\text{gen}}$), and a set of hyper-parameters. The hyper-parameters are: the number of personas generated at each iteration ($B$), the maximum number of iterations ($I$), the generation model ($M_{gen}$), the prediction model ($M_{pred}$), the feedback model ($M_{fb}$), and the generation temperature ($\tau$).

\name's optimization operates as an iterative search, inspired by evolutionary algorithms and LLM-driven optimization approaches~\cite{yang2024large, pryzant-etal-2023, Promptbreeder}. At each iteration, it generates a population of personas, chooses the best one, and provides correction feedback for the subsequent iterations. The search stops when the maximum number of iterations is reached or the best persona shows no more improvements. 
At the first iteration, the generation model is prompted with the persona template, the set of questions ($\mathbf{q}^{\text{gen}}$) and responses ($\mathbf{r}^{\text{gen}}$), and instructions to generate the persona. In this iteration, the model is prompted $B$ times with high temperature to promote diversity in persona generation. The output is a population of $B$ initial personas that follow the persona template and compress the individual questions and responses. We opted to use a higher temperature rather than prompting the model explicitly to diversify the persona generation, as explicit diversity instructions might lead to contrived or superficial variations. 

Given the population of personas, the next step is to evaluate each one of them on the $\mathbf{r}^{\text{gen}}$ set. The procedure prompts a prediction model ($M_{pred}$) with the persona, each question from ($\mathbf{q}^{\text{gen}}$), the answer range, and the instructions to predict the answer. It keeps the persona with the highest accuracy ($acc_{best}$) for the next generation, as the most representative of the user's $\mathbf{r}^{\text{gen}}$. Next, a feedback model ($M_{fb}$) analyzes the best persona at the current iteration for conciseness, generalization, and predictiveness. In particular, it takes the predicted answers from this persona ($\hat{\mathbf{r}}_{best}$) and compares them with the real answers. Then it constructs feedback, which includes necessary corrections to be made to the persona based on wrong predictions, suggested alignment with the template, and improvements for conciseness and generalization.

The next $I-1$ iterations proceed similarly. The generation model ($M_{gen}$) is prompted with the best persona from the prior iteration and the feedback from the feedback model. The generation model produces a batch $B$ of corrected personas using a high temperature to induce diversity. These corrected personas undergo the same process of prediction, selection, feedback, and correction. The search stops at the earlier of two conditions: the maximum number of iterations $I$ has been reached or when the best persona at an iteration matches all the responses from $\mathbf{r}^{\text{gen}}$.

\subsection{Persona-based Prediction} 

\name applies the generated persona to answer a given question on behalf of the individual. This task prompts a prediction model ($M_{pred}$) with three components: the individual's generated persona, the unseen question from ($\mathbf{q}^{\text{eval}}$), and instructions to simulate the decision. Because \name can generate multiple personas for a single user (e.g., one per persona template), it must determine which of them best encapsulates the user's reasoning.

To solve this, \name optionally employs a post-hoc calibration step. Akin to post-hoc adaptation~\cite{krsteski2025, wang2025efficient}, which estimates bias terms on held-out responses to correct statistical properties, we leverage LLMs as a personalized selection mechanism. We designate a small ``calibration set'' from $\mathcal{Q}_{\text{eval}}$. \name evaluates each generated persona on these calibration questions and selects the representation yielding the highest individual-level accuracy ($acc$). The selected persona operates as the final, validated privacy profile for that user. We evaluate the impact of this optional calibration in Section~\ref{sec:evaluation} on simulating privacy decision-making on unseen questions.

%% file: content/41_opt_alog.tex
\begin{algorithm}
\caption{Persona Optimization Procedure}
\label{alg:persona_opt}
\small
\begin{algorithmic}[1]
    \State \textbf{Input:} Persona Template $P_{temp}$, Questions $\mathbf{q}^{\text{gen}}$, User Responses $\mathbf{r}^{\text{gen}}$, Hyper-parameters $(B, I, M_{gen}, M_{pred}, M_{fb}, \tau)$
    \State \textbf{Output:} Optimized Persona $P^*$
    
    \State $P_{best} \gets \text{null}$
    \State $acc_{best} \gets 0$
    \State $F \gets \text{null}$ \Comment{Feedback buffer}

    \For{$i = 1$ \textbf{to} $I$}
        \State \textbf{Step 1: Population Generation}
        \If{$i = 1$}
            \State $\mathcal{P}_i \gets \text{Generate } B \text{ personas from } M_{gen}(P_{temp}| \mathbf{q}^{\text{gen}} | \mathbf{r}^{\text{gen}}, \tau)$
        \Else
            \State $\mathcal{P}_i \gets \text{Generate } B \text{ personas from } M_{gen}(P_{best}|F, \tau)$
        \EndIf
        
        \State \textbf{Step 2: Evaluation and Selection}
        \For{\textbf{each} $P \in \mathcal{P}_i$}
            \State $\hat{\mathbf{r}} \gets \text{Predict answers for } \mathbf{q}^{\text{gen}} \text{ using } M_{pred} \text{ with } P$
            \State $acc \gets \text{Calculate accuracy of } \hat{\mathbf{r}} \text{ against } \mathbf{r}^{\text{gen}}$
            \If{$acc > acc_{best}$}
                \State $acc_{best} \gets acc$
                \State $P_{best} \gets P$
                \State $\hat{\mathbf{r}}_{best} \gets \hat{\mathbf{r}}$
            \EndIf
        \EndFor
        
        \State \textbf{Step 3: Termination Check}
        \If{$acc_{best} = 1.0$ \textbf{or} $i = I$}
            \State \Return $P_{best}$
        \EndIf
        
        \State \textbf{Step 4: Feedback Generation}
        \State $F \gets M_{fb}(P_{best}, \hat{\mathbf{r}}_{best}, \mathbf{r}^{\text{gen}})$ \Comment{Evaluate Conciseness, Generalization, Predictiveness}
    \EndFor
    \State \Return $P_{best}$
\end{algorithmic}
\end{algorithm}

%% file: content/50_eval.tex
\section{Evaluation}
\label{sec:evaluation}

We evaluate how well \name addresses our research questions, and we examine the impact of our design choices. Specifically, our evaluation includes four sets of experiments:

\begin{itemize} 
\item \textbf{Predictiveness and Representation:} Section~\ref{sec:predictiveness} assesses  RQ1 and RQ2 by examining how well text-based personas can reconstruct population-level distributions and predict individual-level privacy decisions. 

\item \textbf{Influence of Attitude and Behavioral Questions:} Section~\ref{sec:attitude_behavior} studies the impact of generating personas based on  attitude versus behavioral questions. We examine if answers to attitude questions can predict behaviors, or if only answers to behavioral questions can do so. 

\item \textbf{Influence of Privacy Behavior Theories:} Section~\ref{sec:eval:theory} quantifies how guiding the persona generation using theoretical frameworks, such as Privacy Calculus or Bounded Rationality, impacts overall predictive accuracy.

\item \textbf{Cross-Study Applicability:} Section~\ref{sec:cross_study} addresses RQ3 by quantifying how well personas synthesized from one study simulate response distributions in an independent study.

\end{itemize}

\begin{table*}[t]
    \centering
    \caption{High-level overview of the five datasets used to evaluate \name.}
    \label{tab:dataset_overview}
    \renewcommand{\arraystretch}{1.05}
    \begin{tabular}{@{} l l l r l c l @{}}
    \toprule
    \textbf{Dataset} & \textbf{Data Source} & \textbf{Date} & \textbf{Size ($N$)} & \textbf{Target Population} & \textbf{\# Qs/respondent} & \textbf{Primary Focus} \\
    \midrule
    \textbf{SPA}~\cite{spa_dataset} & Prolific & Pre-2021 & 1,737 & UK & 160 & Smart assistant sharing \\
    \textbf{Pew PP1}~\cite{pew_wpp1} & Pew ATP & Jan 2014 & 607 & US adults & 51 & General privacy attitudes \\
    \textbf{Pew W49}~\cite{pew_w49} & Pew ATP & Jun 2019 & 4,272 & US adults & 74 & Data sharing trade-offs \\
    \textbf{Pew W127}~\cite{pew_w127} & Pew ATP & May 2023 & 5,101 & US adults & 131 & Tech \& privacy behaviors \\
    \textbf{CAuthN}~\cite{cauthn_dataset} & Prolific & Sep--Oct 2022 & 830 & US Internet users & 83 & Biometric authentication \\
    \bottomrule
    \end{tabular}
\end{table*}

\subsection{Datasets}
\label{sec:datasets}

We utilize five privacy-centric datasets (Table~\ref{tab:dataset_overview}) to carry out the above evaluations. We choose three types of surveys to represent different granularities in the measured privacy decisions. Three datasets are privacy panels from Pew Research that measure general privacy attitudes and behaviors. The fourth dataset from Dehling et al. measures users' willingness to disclose biometric data for continuous authentication~\cite{cauthn_dataset}. The fifth dataset measures whether individuals accept a smart personal assistant sharing specific data types with third-party entities for specific purposes. We describe these datasets briefly here, and defer the full description to Appendix~\ref{sec:app:datasets}. 

\paragraph*{SPA Dataset~\cite{spa_dataset}} 
Abdi et al. published a dataset about privacy norms for smart home personal assistants (SPAs)~\cite{spa_dataset} comprising 1,737 responses. The survey measures participants' comfort with a set of 24 data-sharing scenarios, alongside their privacy and security attitudes. The scenarios cover varied data types, recipients, purposes, and conditions. When analyzing this dataset, we partition the responses into two groups: behaviors (i.e., data sharing scenarios mapped to acceptable/unacceptable), and privacy and security attitudes. We discard neutral responses.

\paragraph*{Pew PP1~\cite{pew_wpp1}}
The Pew Research Center conducted this study in 2014 with 607 US adults to measure their perceptions of privacy, surveillance, and data management. This dataset covers perceived importance of controlling personal information, concern over government and corporate monitoring, and the perceived sensitivity of various data types.

\paragraph*{Pew W49~\cite{pew_w49}}
Collected by Pew in 2019, this dataset focuses on the trade-offs inherent in the modern data ecosystem. It includes responses from 4,272 U.S. adults. The study covers perceived risks versus benefits of data collection, concerns regarding data usage, sense of control over data, and understanding of privacy policies. It also includes hypothetical scenarios where respondents rated the acceptability of data sharing in exchange for specific services or societal benefits.

\paragraph*{Pew W127~\cite{pew_w127}}
Collected by Pew in 2023, this study includes responses from a nationally representative sample of 5,101 U.S. adults. The survey covers general concerns regarding government and corporate data collection, perceived control over personal data, individual digital privacy behaviors (e.g., use of password managers), and specific attitudes toward emerging technologies, such as AI and residential cameras.

\paragraph*{CAuthN Study Dataset~\cite{cauthn_dataset}}
Dehling et al. studied individuals' perceptions around continuous biometric authentication. They recruited 830 US Internet users in 2022. The study covers willingness to continuously share biometric data, belief in the efficacy of biometric authentication, perceived privacy and security risks, and perceived trust in the service provider. It also includes standard privacy and security attitude scales.

\subsection{Experiments}
\label{sec:experiments}

\noindent
We perform four sets of experiments as detailed here.

\subsubsection{Predictiveness and Representation}
\label{sec:predictiveness}
The first set of experiments investigates whether synthetic personas can predict responses to unseen questions drawn from the same surveys at both the individual and population levels.

\paragraph{Experiment Setup}

We represent each study as the dataset matrix defined in Section~\ref{sec:problem}. The first set of experiments does not distinguish between attitude and behavioral questions. We randomly split the response set for each user into two subsets: a generation set (80\%) and an evaluation set (20\%). 
There is no overlap between the questions used to generate the persona and those used to evaluate it. We use Gemini 3.0 Flash (temperature = 1.5) to synthesize narratives and Gemini 2.5 Flash-Lite (temperature = 0) for prediction. The optimization parameters are $B=5$ personas per iteration and $I=3$ iterations. In this evaluation, we use the basic persona generation template (Appendix~\ref{appendix:generation_prompts}) to synthesize, and the basic prediction template (Appendix~\ref{appendix:prediction_prompts}) to apply the personas.

To compute population-level metrics, we aggregate the simulated answers per unique evaluation question to construct a predicted distribution. We then compute the TVComplement between the predicted distribution and the ground-truth distribution for the specific subset of respondents who answered the question. We report the average across all questions. Finally, we evaluate the text-based personas against two prompting strategies: \textit{raw examples} and \textit{internal LLM baseline}. The raw examples strategy refers to using the survey question-answer sets verbatim as the privacy persona. The internal LLM baseline approach instructs the LLM to apply standard privacy norms without any personalization (see Appendix~\ref{appendix:prediction_prompts} for exact prompts). We report the individual-level accuracy ($acc_S$) and population-level metric ($1-TVD$) as defined in Section~\ref{sec:problem}.

\begin{figure}[t]
\centering
\includegraphics[width=\linewidth,keepaspectratio]{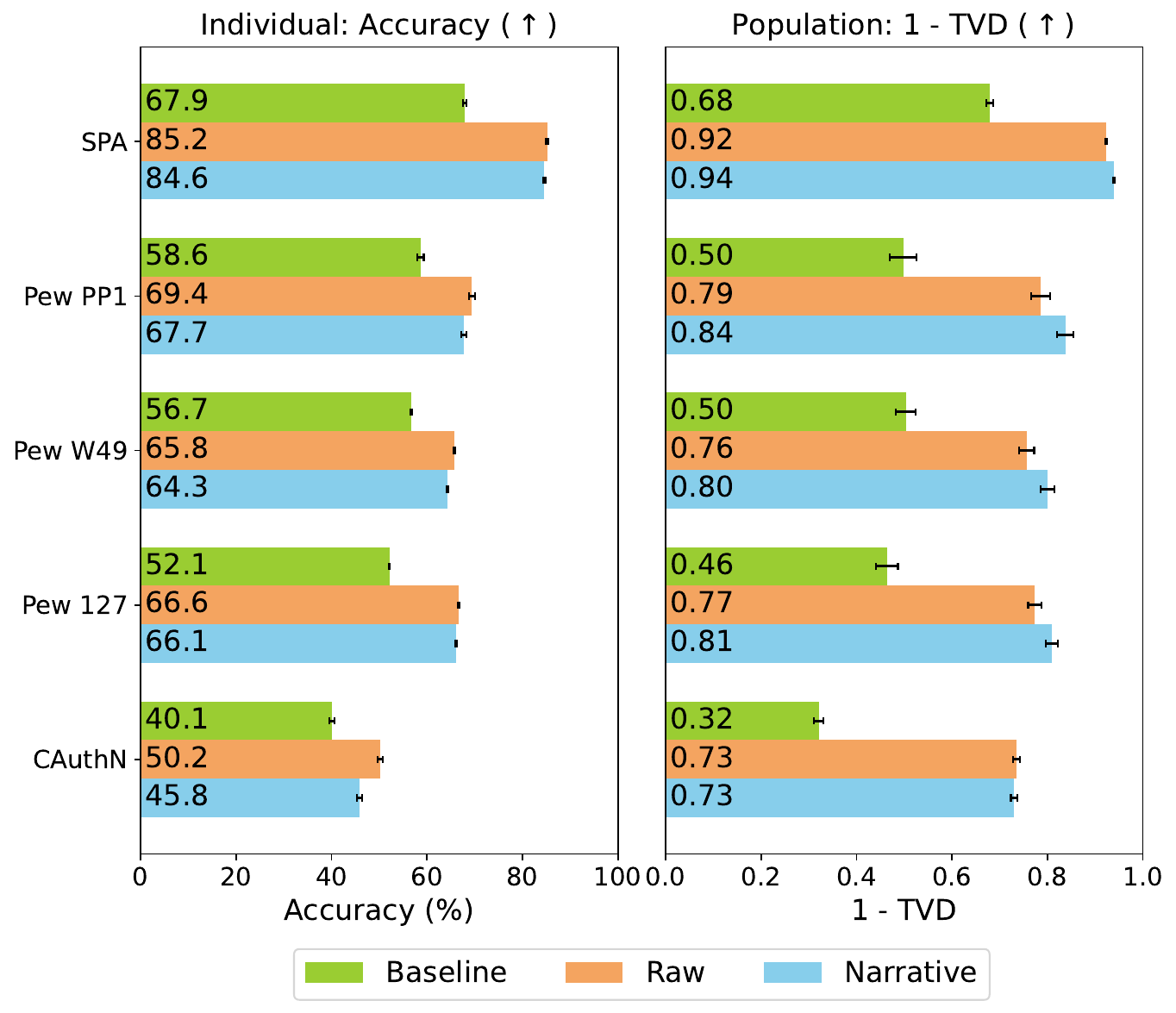}
\caption{Individual- and population-level performance of text-based personas using 80:20 generation/evaluation split. Error bars show 95\% confidence intervals obtained via bootstrapping. We report performance of summarized persona derived from question-answer pairs (Narrative), against the original question-answer pairs (Raw) and the LLM's privacy baseline without personalization (Baseline). %
}
\label{fig:set1_main_result}
\end{figure}

\paragraph{Results}
Fig.~\ref{fig:set1_main_result} shows the individual-level and population-level performance of the text-based personas on the five datasets. Error bars represent 95\% confidence intervals obtained through bootstrapping the results 1,000 times with replacement. We highlight key takeaways while deferring detailed metrics to Appendix~\ref{appendix:detailed-results}.

\noindent\textbf{Text-based personas retain predictive power.} First, the performance of the text-based persona tracks that of the raw examples across all datasets and metrics. This tight alignment highlights the effectiveness of \name in distilling the generation set. Rather than losing important information about user decisions during the optimization, the text-based persona successfully captures the user's core privacy preferences from the raw question-answer pairs.

\noindent\textbf{Personalization is strictly necessary.} Second, both raw examples and text-based personas perform significantly better than the internal LLM baseline on both individual- and population-level metrics. Prompting LLMs to answer privacy-related questions without grounding fails to match the real distribution of responses~\cite{li2026llmagentssimulateenduser}. Grounding, either through exact data or a summarized persona, is essential for accurate predictions.

\noindent\textbf{Predictability varies among respondents in the same dataset.}  Third, at the individual-level, we observe significant variance in the prediction accuracy for each dataset with the standard deviation varying between 10\% and 15\%. Still, \name accurately represents specific subsets of users: more than 40\% of respondents achieve $>90\%$ matching accuracy on the SPA dataset, and more than 30\% of respondents achieve $>70\%$ accuracy in the Pew W49 and PP1 datasets. We attribute this variance to several factors related to inconsistencies in responses~\cite{krsteski2025}, decision-making, views of institutions, tech savviness, and assessments of data sensitivity. Section~\ref{sec:qual_analysis} investigates this aspect further through qualitative analysis of predictive and non-predictive personas.

\noindent\textbf{Predictability varies significantly across datasets.} 
Fourth, we observe a noticeable variance in absolute prediction accuracy across the five datasets. We attribute this variance to the dataset structure in terms of number of questions and answer granularity, where datasets with highly repetitive structures enable stronger extrapolation. For instance, the SPA dataset contains $\sim$144 data sharing questions per respondent that follow the same structure, allowing the model to achieve high predictability. Conversely, the CAuthN dataset involves predicting continuous values on a 1--100 scale, resulting in poorer baseline accuracy. On the Pew datasets, the models achieve $\sim$65\% to 70\% average individual accuracy on 5- and 7-point Likert scales, where random baseline accuracy would be 14-20\%.

\noindent\textbf{Personas accurately reconstruct population distributions.} Fifth, \name demonstrates high fidelity at the population-level. As shown in the right plot of Fig.~\ref{fig:set1_main_result}, the text-based personas consistently yield the highest TVComplement ($1 - \text{TVD}$) across all datasets, outperforming raw examples by 0.01-0.05 - a consistent improvement despite using 80-95\% fewer tokens.
This result indicates that grounded personas accurately mimic population-level aggregated responses on unseen questions, effectively smoothing out some of the individual-level noise inherent in survey data.

\noindent\textbf{Nuanced privacy preferences are harder to predict.} Sixth, like the individual-level predictability, population-level performance varies across specific evaluation questions. To illustrate why some questions are inherently easier to predict than others, we highlight two examples from the SPA dataset. One question asked participants if sharing smart thermostat history with an entertainment skill (e.g., Star Wars quotes) was acceptable. This proved easy to predict: the population-level metric (TVComplement) was perfect (\name correctly estimated the distribution at 47/54 unacceptable and 7/54 acceptable), and individual-level accuracy was 50 of 54. This high accuracy likely stems from a clear lack of functional necessity and an obvious privacy risk of revealing whether someone is home or not. In contrast, a more difficult question asked if sharing sleeping hours with a health and fitness skill was acceptable. For this scenario, TVComplement dropped to 0.75 and individual accuracy dropped to 41 of 58 responses. This question captures a subjective trade-off: some users gladly share sleep data for personalized health analysis, while others consider it too private regardless of the benefit. These examples demonstrate that predictive models more easily internalize clear-cut privacy violations than nuanced, subjective trade-offs.

\noindent\textbf{Text-based personas are token-efficient.} 
Finally, Table~\ref{tab:token_counts} reports the number of tokens for both the raw examples and \name's personas across the datasets. The token count is calculated per individual using the Gemini 3.0 tokenizer, then averaged across all individuals per dataset. As shown in the table, \name generates highly concise personas, particularly when the generation set is large. For instance, the SPA dataset contains over 100 question-answer pairs in its generation set, yet \name's personas are approximately $20\times$ smaller in token count compared to the raw examples. 
We simulate the cumulative costs of both approaches over a sequence of 1,000 questions to assess long-term cost-effectiveness. Even when factoring in the initial generation cost and periodic updates to account for preference drift (every 10 interactions), \name becomes more cost-effective than maintaining full raw histories within approximately 94 to 164 interactions across datasets. Beyond the break-even point, the cumulative cost of applying raw examples grows quadratically while that of personas grows linearly. This scaling ensures efficiency, while concise personas improve auditability and portability without sacrificing predictiveness or representativeness.

\begin{table}[t]
\centering
\begin{tabular}{lcccc}
\toprule
\textbf{Category} & \textbf{Raw} & \textbf{Narrative} & \textbf{\% Reduction} &\textbf{Break-Even Point}\\
\midrule
SPA & 8920 & 431 & 95\% & 96 questions\\
Pew PP1 & 1738 & 296 & 82\% & 95 questions\\
Pew W49 & 2291 & 399 & 82\% & 118 questions\\
Pew W127 & 3243 & 453 & 86\% & 165 questions\\
CAuthN & 3006 & 395 & 86\% & 109 questions\\
\bottomrule
\end{tabular}
\caption{The average token count comparison between raw examples and the personas from the basic template. The narrative personas reduce token count by 82\%-95\% compared to the baseline.}
\label{tab:token_counts}
\end{table}

\subsubsection{Influence of Attitude and Behavioral Questions}
\label{sec:attitude_behavior}
In the second set of experiments, we delve deeper into whether attitude questions can predict responses to behavioral questions. 

\begin{figure}[t]
\centering
\includegraphics[width=\linewidth,keepaspectratio]{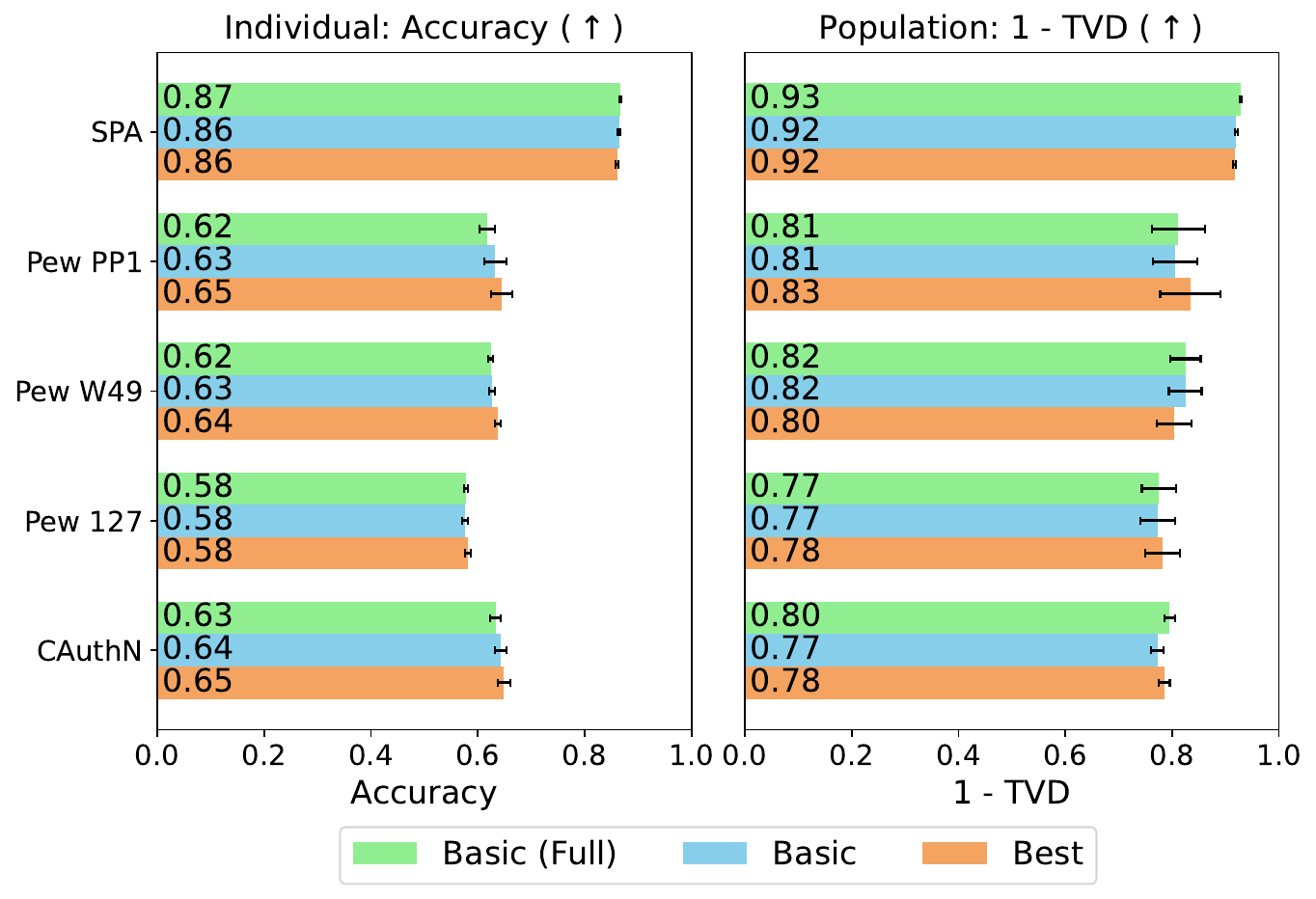}
\caption{Performance when generating from behavioral questions and evaluating on behavioral questions, across prompt templates. Error bars are obtained via bootstrapping.}
\label{fig:split_bb_metrics_plot_aggregate}
\end{figure}

\begin{figure}[t]
\centering
\includegraphics[width=\linewidth,keepaspectratio]{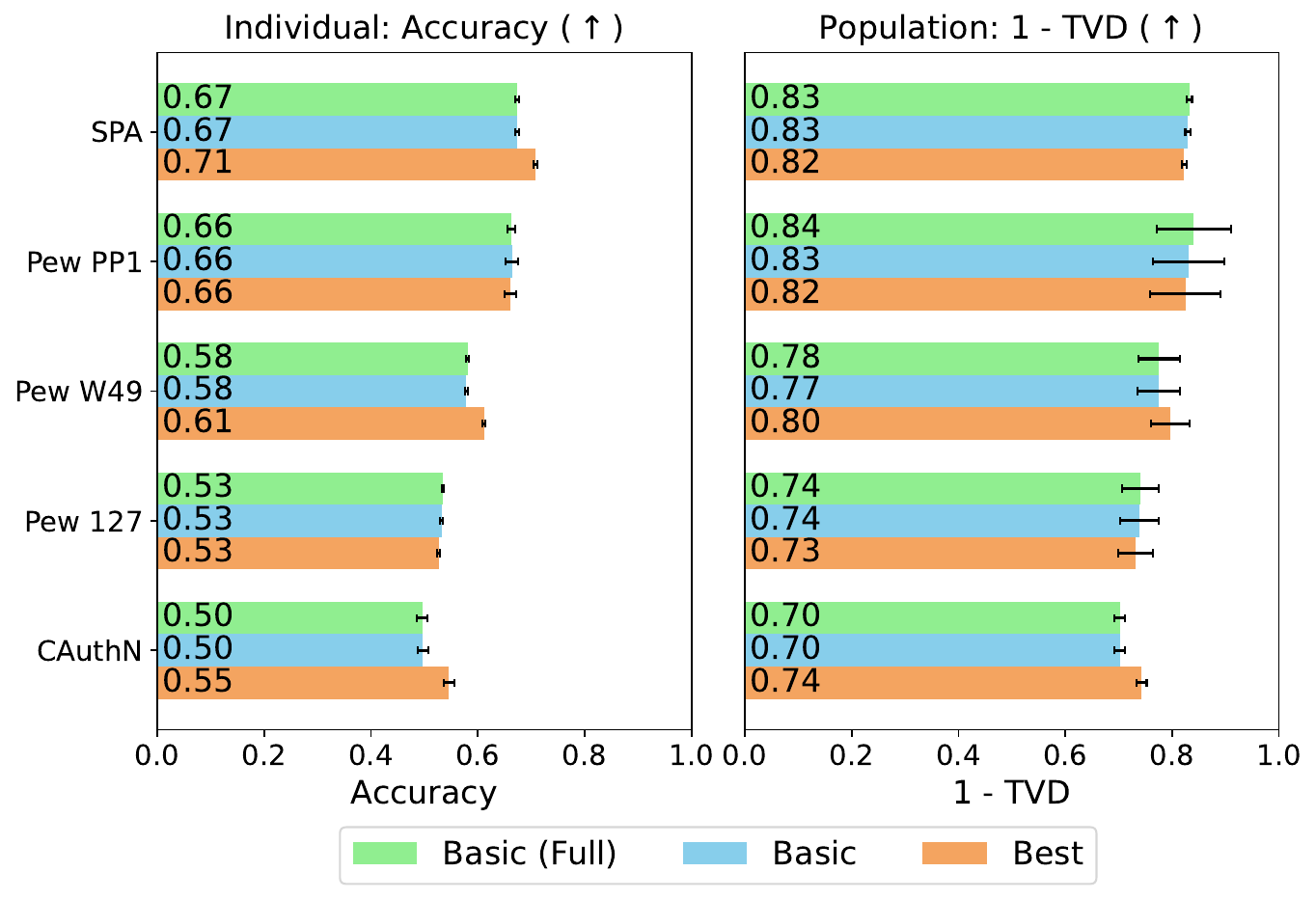}
\caption{Performance when generating from attitude questions and evaluating on behavioral questions, across prompt templates. Error bars are obtained via bootstrapping. }
\label{fig:split_ab_metrics_plot_aggregate}
\end{figure}

\paragraph{Experiment Setup}
We split the questions in each dataset into attitude and behavioral domains. Attitude questions quantify a respondent's  perceptions, values, and concerns regarding privacy, such as the perceived sensitivity of information, level of concern regarding surveillance, or trust in institutional data handling. Behavioral questions measure self-reported past actions or intended decisions, such as the adoption of privacy-enhancing tools or the willingness to share specific data types. We manually label each question as attitude or behavioral. Questions not falling into either category are discarded, such as demographic questions. We consider two experimental settings: (1) splitting the behavioral questions between generation (80\%) and evaluation (20\%), and (2) using attitude questions for generation and behavioral questions for evaluation. We generate personas in a single iteration ($I=1$).

\paragraph{Results}
Fig.~\ref{fig:split_bb_metrics_plot_aggregate} and Fig.~\ref{fig:split_ab_metrics_plot_aggregate} show how the text-based personas predict individual responses under these two experimental conditions. The detailed results across templates are found in Appendix~\ref{appendix:detailed-results}. Our analysis here focuses on the performance of the Basic template (labeled as ``Basic (Full)''), where we highlight two main takeaways.

\noindent\textbf{Behavioral questions are stronger predictors of other behavioral decisions.} As shown in Fig.~\ref{fig:split_bb_metrics_plot_aggregate}, generating personas directly from answers to behavioral questions provides a reliable basis for predicting responses to unseen behavioral questions, a finding consistent with prior studies~\cite{flemings2025}. When comparing performance on a per-dataset basis, isolating behavioral questions for generation consistently yields better accuracy than the baseline case where domains are randomly mixed (detailed previously in Fig.~\ref{fig:set1_main_result}). For example, on the SPA dataset, personas built strictly from behavioral questions achieve up to 87\% individual accuracy, compared to the 85\% achieved with a random split.

\noindent\textbf{General privacy attitudes are weak predictors of concrete behaviors.} Conversely, models struggle to predict answers to concrete behavioral privacy questions when personas are derived solely from general attitude surveys. By contrasting these two experimental settings, we find that attitude-based generation (Fig.~\ref{fig:split_ab_metrics_plot_aggregate}) results in lower individual- and population-level metrics than behavior-based generation (Fig.~\ref{fig:split_bb_metrics_plot_aggregate}). Although confined to survey settings, this predictive gap may reflect the ``privacy paradox,'' where stated concerns frequently misalign with practical data-sharing realities. When relying on attitude-based rather than behavior-based generation, individual accuracy drops significantly: falling from 87\% to 67\% on the SPA dataset, and from 63\% to 50\% on CAuthN. Population-level TVComplement simultaneously decreases across the board (e.g., TVComplement on CAuthN degrades from $0.8$ to $0.7$).

\subsubsection{Influence of Privacy Behavior Theories}
\label{sec:eval:theory}
Next, we assess the impact of privacy theories on \name's performance.

\paragraph{Experiment Setup}
We employ three additional persona templates inspired by established privacy theories: Privacy Calculus, Bounded Rationality, and Protection Motivation Theory (PMT) (described in Section~\ref{sec:templates} and Appendix~\ref{appendix:privacy_theories}). We keep the same generation and prediction parameters as Section~\ref{sec:predictiveness}. Our goal is to measure whether framing the prompt around how individuals make data-sharing decisions impacts the predictive accuracy of the generated personas. We use Gemini 3.0 Flash to generate privacy persona narratives based on half of an individual's responses, applying one of the four templates (Basic, Bounded, Calculus, PMT; the generation prompts are in Appendix~\ref{appendix:generation_prompts}).  We then evaluate these persona narratives on the remaining questions to compute population-level metrics and individual-level accuracy. When generating the personas, we set $I=1$, as our analysis in Fig.~\ref{fig:tvd_iterations} from Appendix~\ref{appendix:detailed-results-iteration-steps} shows that additional optimization steps are unlikely to result in better-performing personas.

\paragraph{Results}
 We highlight three main findings across the datasets. A detailed set of results is in Appendix~\ref{appendix:detailed-results}.

\noindent\textbf{Aggregate performance is robust to theoretical framing.} Table~\ref{tab:narrative_template_choice_performance_tvd} reports the TVComplement metric across the prompt variants, demonstrating that population-level performance remains highly stable (e.g., $1-TVD$ is between $0.80$ and $0.81$ on Pew W127 across all prompt types). This stability confirms that, in aggregate, persona predictiveness stems primarily from grounding the LLM in historical user data rather than the framing of the generation prompt. Importantly, while framing does not affect accuracy, it provides a standardized, modular format that supports auditability and editability.

\begin{figure}[t]
\centering
\includegraphics[width=\linewidth,keepaspectratio]{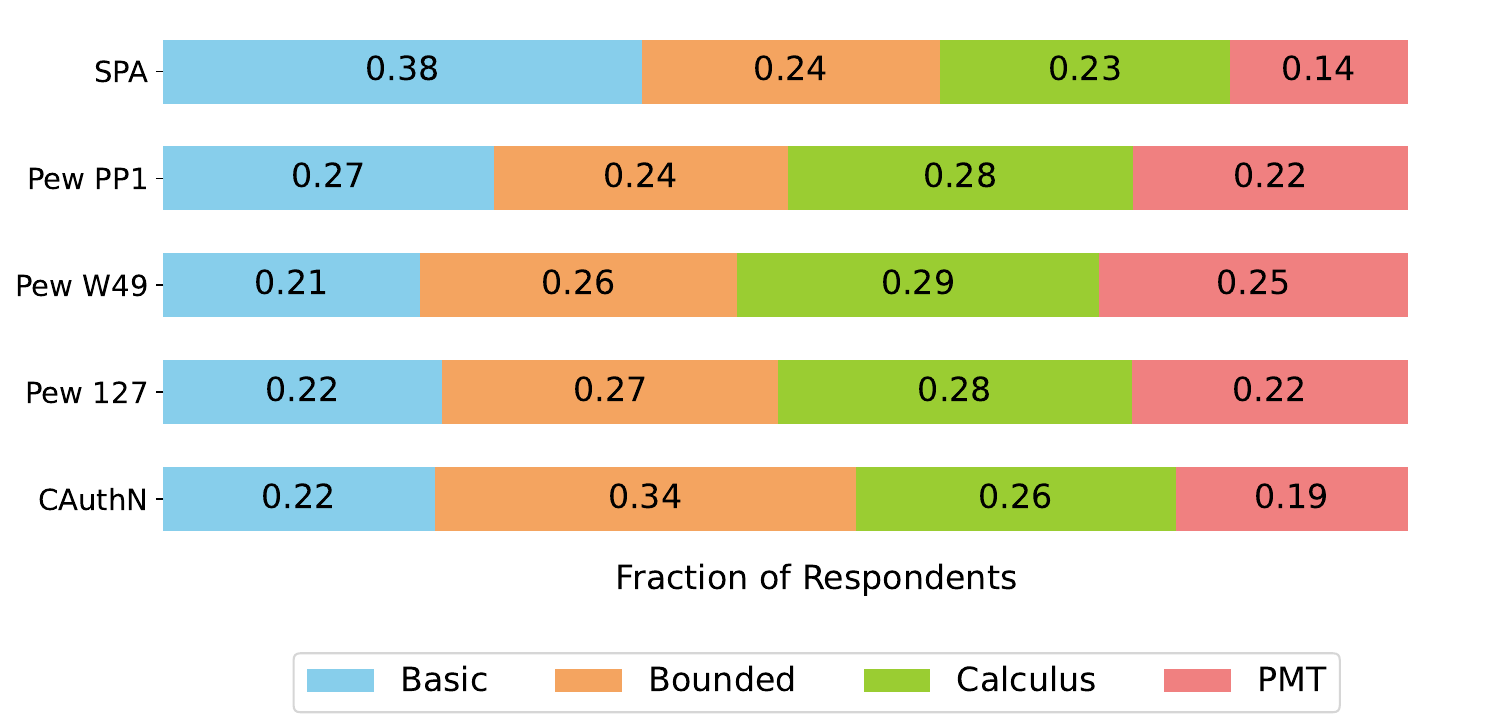}
\caption{The fraction of individuals in the five datasets for which each privacy theory template yields the highest predictive accuracy.}
\label{fig:split_0.2_metrics_best_synthesizer}
\end{figure}

\noindent\textbf{No single privacy theory dominates globally.} While aggregate performance is stable, individual-level variance is significant. Fig.~\ref{fig:split_0.2_metrics_best_synthesizer} shows the fraction of individuals for whom each theory yielded the highest predictive accuracy. No single template universally captures everyone's behavior. For instance, on the SPA dataset, the Basic template performs the best for 38\% of users, but Bounded Rationality and Privacy Calculus best fit 24\% and 23\% of users, respectively. This fragmentation suggests that different individuals naturally align with different theoretical heuristics when evaluating privacy decisions.

\noindent\textbf{Dynamically matching templates to respondents can unlock higher accuracy.} Leveraging this fragmented alignment, we find a measurable boost in performance when dynamically selecting the optimal theoretical template per respondent. Towards that end, we further break down the evaluation set in both settings into disjoint calibration and test subsets. We use the calibration set for each user to dynamically select the template that yields the highest performance. We then report the accuracy of this template on the test subset. The bars titled ``Best'' in both Fig.~\ref{fig:split_bb_metrics_plot_aggregate} and Fig.~\ref{fig:split_ab_metrics_plot_aggregate} refer to this dynamic selection scheme, while the bar titled ``Basic'' refers to the performance of the Basic template on the test subset. In the behavior-to-behavior setting, we observe no difference in performance because the number of evaluation questions was already small (Table~\ref{tab:question_count_bb} in Appendix~\ref{appendix:detailed-results-question-count}). However, we see statistically significant gains on some datasets in the attitude-to-behavior setting, where the number of evaluation questions is large enough. For example, Fig.~\ref{fig:split_ab_metrics_plot_aggregate} shows that performance for the CAuthN dataset increases for both individual- and population-level metrics compared to the ``Basic'' persona. This result reveals a practical computational trade-off: investing generation resources to synthesize and evaluate multiple theoretical personas for a user ultimately provides a direct pathway to superior predictive accuracy over time.

\begin{table}[h]
    \centering
\resizebox{\columnwidth}{!}{%
\begin{tabular}{  c  c c c c}%
\toprule
 \textbf{Dataset} & \textbf{Basic} & \textbf{Bounded} & \textbf{Calculus} & \textbf{PMT} \\ 
 \midrule
SPA & 0.94 ($\pm$0.00) & 0.93 ($\pm$0.00) & 0.94 ($\pm$0.00) & 0.93 ($\pm$0.00) \\
Pew PP1 & 0.84 ($\pm$0.02) & 0.84 ($\pm$0.02) & 0.82 ($\pm$0.02) & 0.83 ($\pm$0.02) \\
Pew W49 & 0.80 ($\pm$0.01) & 0.80 ($\pm$0.01) & 0.79 ($\pm$0.01) & 0.80 ($\pm$0.01) \\
Pew W127 & 0.81 ($\pm$0.01) & 0.81 ($\pm$0.01) & 0.80 ($\pm$0.01) & 0.80 ($\pm$0.01) \\
CAuthN & 0.73 ($\pm$0.01) & 0.73 ($\pm$0.01) & 0.73 ($\pm$0.01) & 0.73 ($\pm$0.01) \\
 \bottomrule
\end{tabular}
}
\caption{Impact of the persona-generating prompt on the population-level metric $1-TVD$ ($\uparrow$), on the five datasets. The performance is similar across prompt variants for each dataset, suggesting that the high predictiveness of the personas does not stem from the choice of prompt.}
\label{tab:narrative_template_choice_performance_tvd}
\end{table}

\subsubsection{Cross Study Applicability}
\label{sec:cross_study}
The final set of experiments evaluates the cross-study applicability of \name by assessing whether text-based personas from one study can simulate the privacy decisions in a different study.

\paragraph{Experiment Setup}

We evaluate three cross-study scenarios, by generating personas from one study's questions, selecting highly performing ones, and applying them to questions from another study. We focus on the three Pew datasets (PP1--2014, W49--2019, and W127--2023) because they are representative of the U.S. adult population by design. The PP1 participants are drawn from the KnowledgePanel~\cite{ipsosKnowledgePanel} (formerly operated by GfK) and the W49/W127 participants are drawn from the American Trends Panel~\cite{pewATP}. The KnowledgePanel (now operated by Ipsos) recruits members using address-based sampling to ensure representation of the U.S. population, including households that were previously offline. Similarly, the American Trends Panel is Pew Research Center's representative panel of U.S. adults. To mitigate distribution bias from lack of responses and coverage, both platforms apply statistical weighting to their final datasets. The weighted datasets align with U.S. Census benchmarks for demographic characteristics such as age, gender, race, and education.

\noindent
\newline\textbf{Scenarios:}
We evaluate three scenarios of cross-study applicability. In the first scenario (Table~\ref{tab:cross_study_results_att_beh}), we divide each survey's questions into attitude and behavioral domains (same splits as Section~\ref{sec:attitude_behavior}). We generate personas based on attitude questions of an earlier study (source) and apply them to behavioral questions of a later study (target). We compare the performance of the transferred personas to an in-study baseline, where personas are generated on attitude questions and applied to behavioral questions of the same target study. 

In the second scenario (Table~\ref{tab:cross_study_results_beh_beh}), we randomly split the behavioral questions into generation (80\%) and evaluation sets (20\%). We generate personas based on the generation set of the source study and apply them to the evaluation set of the target study. This question split matches that of Section~\ref{sec:attitude_behavior}. The in-study baseline is where the generation and evaluation sets belong to the behavioral questions of the same target study.

 In the third scenario (Table~\ref{tab:cross_study_synth}), we evaluate synthetic personas derived solely from demographic attributes. We generate these synthetic personas based on demographic questions from the Pew W127 study and apply them to the behavioral questions in the target studies (W49, W127, and PP1). We utilize demographic attributes that include: location (census division and urban/rural identification), basic demographics (age, gender, race, and ethnicity), background and origins (birthplace, citizenship, and years lived in the U.S.), socioeconomic indicators (family income and marital status), and cultural or political markers (religion, political party affiliation, ideology, and voter registration status). Because all three studies mirror the same underlying distribution, using the demographic questions from one of them is sufficient to yield a representative synthetic panel. We compare the baseline predictive performance of these unrefined personas (raw) against those from a single optimization iteration (iteration 1).

\noindent
\newline\textbf{Persona Selection and Evaluation:} Given the generation set of questions from a source study, we generate candidate personas using the basic template. To measure population-level metrics across studies, we use a Cartesian matrix approach. Specifically, we use each selected persona to generate answers to all unique questions in the evaluation set of the target dataset. We then compare the distribution of the simulated answers to that of the real answers on the target dataset. Otherwise, we use the same generation and prediction parameters as in Section~\ref{sec:predictiveness}. Finally, to ensure that our evaluation reflects population-level distributions, we incorporate survey response weights into our metric computations. We provide the details of this weighted evaluation in Appendix~\ref{app:more_metrics}.

\begin{table}[t]
\centering
\renewcommand{\arraystretch}{0.95}
\begin{tabular}{@{} l c c @{}}
\toprule
\textbf{Scenario (Att$\rightarrow$Beh)} & \textbf{$1-TVD$ ($\uparrow$)} & In-Study Baseline \\
\midrule
PP1 $\rightarrow$ W49 & $0.772 \pm 0.030$ & $0.782 \pm 0.039$ \\
PP1 $\rightarrow$ W127 & $0.808 \pm 0.028$ & $0.744 \pm 0.035$ \\
W49 $\rightarrow$ W127 & $0.729 \pm 0.032$ & $0.744 \pm 0.034$ \\
\bottomrule
\end{tabular}

\caption{Population-level metrics measuring the cross-study applicability of personas (Attitude to Behavior). Higher values indicate closer alignment with the real distribution of the target dataset.}
\label{tab:cross_study_results_att_beh}
\end{table}

\begin{table}[t]
\centering
\renewcommand{\arraystretch}{0.95}
\begin{tabular}{@{} l c c @{}}
\toprule
\textbf{Scenario (Beh$\rightarrow$Beh)} & \textbf{$1-TVD$ ($\uparrow$)} & In-Study Baseline \\
\midrule
PP1 $\rightarrow$ W49 & $0.716 \pm 0.039$ & $0.817 \pm 0.029$ \\
PP1 $\rightarrow$ W127 & $0.676 \pm 0.036$ & $0.780 \pm 0.035$ \\
W49 $\rightarrow$ W127 & $0.782 \pm 0.034$ & $0.780 \pm 0.034$ \\
\bottomrule
\end{tabular}

\caption{Population-level metrics measuring the cross-study applicability of personas (Behavior to Behavior). Higher values indicate closer alignment with the real distribution of the target dataset.}
\label{tab:cross_study_results_beh_beh}
\end{table}

\begin{table}[t]
\centering
\renewcommand{\arraystretch}{0.95}
\begin{tabular}{@{} l c c @{}}
\toprule
\textbf{Target Study} & \textbf{1-TVD (raw)} & \textbf{1-TVD (iteration 1)} \\
\midrule
synthetic $\rightarrow$ W49 & $0.632 \pm 0.047$ & $0.739 \pm 0.027$ \\
synthetic $\rightarrow$ W127 & $0.585 \pm 0.038$ & $0.716 \pm 0.026$ \\
synthetic $\rightarrow$ PP1 & $0.751 \pm 0.065$ & $0.757 \pm 0.062$ \\
\bottomrule
\end{tabular}
\caption{Population-level metrics measuring the effectiveness of demographics only attributes in predicting privacy attitudes. Higher values indicate closer alignment with the target dataset's distribution.}
\label{tab:cross_study_synth}
\end{table}

\paragraph{Results}
Tables~\ref{tab:cross_study_results_att_beh}, \ref{tab:cross_study_results_beh_beh}, and \ref{tab:cross_study_synth} report the population-level transfer performance across our three scenarios. Below are the main takeaways.

\noindent
\textbf{Behavioral trends transfer well across small temporal gaps.} 
When applying personas across populations with small temporal gaps, \name mimics population-level trends. The personas generated from Pew W49 (2019) transfer well to Pew W127 (2023) for both Att$\rightarrow$Beh and Beh$\rightarrow$Beh scenarios. We observe the same trend between Pew PP1 (2014) and Pew W49 (2019). In both scenarios, the population level metrics obtained from cross-study personas match those obtained from the in-study baselines. For instance, in Table~\ref{tab:cross_study_results_beh_beh}, personas generated from Pew W49 (2019) behavioral questions transfer well to Pew W127 (2023) behavioral questions ($1-TVD=0.782$). This performance is comparable to the in-study baseline of personas generated on the Pew W127 dataset ($1-TVD=0.780$). This result shows that generating personas based on recent responses from the same population in a similar domain can simulate current population-level phenomena.

\noindent
\textbf{Longitudinal behavior drift limits long-term behavior-to-behavior transferability.} 
We observe that behavior-grounded personas do not transfer reliably across large temporal gaps, especially from Pew PP1 (2014) to Pew W127 (2023), where performance drops to $1-TVD=0.676$ compared to the $0.780$ baseline (Table~\ref{tab:cross_study_results_beh_beh}). We attribute this discrepancy to the evolution of the privacy landscape, technologies, and data-sharing behaviors over the nine-year gap, rendering the 2014 behavioral personas outdated for modern scenarios. 

\noindent
\textbf{Attitude-based personas exhibit robust long-term cross-study transferability.} 
When applying personas generated from attitude questions to the behavioral domains (Table~\ref{tab:cross_study_results_att_beh}), the older PP1 (2014) dataset serves as a strong source. Personas generated from PP1 attitude questions applied to W127 behavioral questions achieve a $1-TVD$ of $0.808$, outperforming the in-study baseline ($0.744$). This suggests that while specific privacy behaviors drift over time, privacy attitudes can be more stable. However, current behavioral questions remain more predictive as evident from the Beh$\rightarrow$Beh in-study baselines.

\noindent
\textbf{Demographic-only synthetic personas provide a viable baseline but require optimization.} Table~\ref{tab:cross_study_synth} shows the performance of demographic-based personas. Completely synthetic personas, derived solely from demographic attributes (albeit detailed) without prior survey responses, achieve moderate predictive performance (raw $1-TVD$ ranging from $0.585$ to $0.751$). However, applying a single optimization iteration significantly improves their alignment with real distributions (e.g., improving from $0.585$ to $0.716$ on W127). This result highlights that large language models can augment demographic attributes with contextual privacy reasoning, improving their ability to model population-level privacy decisions.

\noindent
\textbf{Reconciling the temporal dynamics of attitudes and behaviors.} On the surface, our cross-study findings might seem inconsistent with Section~\ref{sec:attitude_behavior}, where we found that attitude-based personas underperform behavior-based ones. We interpret this as evidence that attitudes and behaviors operate at different time scales rather than an inconsistency. Behavioral responses are tightly coupled to specific technologies that evolve rapidly, whereas attitudes capture more stable underlying values. Overall, our results show that behavioral questions are better predictors of individual-level user privacy behavior, but their predictiveness decays faster. At the same time, attitude questions are worse at the individual-level, but they are more robust over time. This insight suggests a strategy to construct personas grounded in privacy attitude questions and refresh them periodically with current behavioral questions.

%% file: content/51_qual.tex
\section{Qualitative Analysis of Personas}
\label{sec:qual_analysis}

Predictability varies significantly across personas, with individual accuracy ranging from over 95\% to under 40\%. To identify what features contribute to this variance, we analyzed the top and bottom 60 personas per study across the five studies (600 in total), using an LLM-assisted thematic analysis approach~\cite{lam-lloom-chi2024,patat_2023}. We prompted Gemini 3.1 Pro to perform initial coding, which yielded 54 codes describing characteristics for high- and low-predictive personas. We then validated these codes by manually reviewing 100 personas (20 personas per study). Following this validation, two authors collaboratively conducted a manual thematic analysis~\cite{Braun01012006}, by examining sample personas and iterating on the codes and their definitions. Eventually, they synthesized the initial codes into six high-level themes. We summarize the identified themes below, along with \textit{excerpts} from relevant personas.

{\bf User Decision Style}. Predictable personas employ categorical rules (e.g., “always share with family" or “never share location"). In contrast, unpredictable personas lack clear boundaries and often make decisions that contradict their established logic. \\
\underline{High accuracy:}  \textit{“sharing is acceptable only if the user is granted the explicit right to review or delete the data.”} A persona can be nuanced but predictable as long as the person remains consistent; for example, \textit{“he supports a right to be forgotten for personal embarrassments and media coverage yet believes law enforcement records should remain public.”} \\
\underline{Low accuracy:} \textit{“she is highly willing to share data like fingerprints despite perceiving high risks, but she is unwilling to share voice prints despite recognizing their utility."} This example illustrates the challenge to predict what this persona would do with a new data type (e.g., 3D facial mapping). The persona doesn't exhibit a consistent decision making logic because of sharing high-risk fingerprints but rejecting to share high-utility voice prints.

{\bf Data Sensitivity}. Predictable individuals have clear views about specific data types (e.g., medical, financial). Low accuracy personas describe someone with varying preferences, sometimes sharing a data type while withholding it in similar contexts, and/or violating common sharing norms. \\
\underline{High accuracy:} \textit{“persona refuses to share biometrics”} \\
\underline{Low accuracy:} \textit{``Their personal sharing logic is highly fragmented. They are comfortable sharing health data and camera status with peripheral figures like neighbors and visitors for safety and health monitoring. However, they consistently exclude siblings from accessing health or camera data. Conversely, they share voice history with their siblings and parents, but refuse to share camera status with their partner."}

{\bf Social Trust Hierarchy}. High-accuracy personas follow logical hierarchies (e.g., Family $>$ Friends $>$ Strangers). Unconventional trust hierarchies remain predictable as long as individuals maintain a consistent internal logic. Low-accuracy personas exhibit erratic or contradictory social trust patterns. \\
\underline{High accuracy:} \textit{“sharing device status is acceptable for partners and children, but all other relatives and outsiders are excluded.”} \\
\underline{Low accuracy:} \textit{“they maintain an idiosyncratic boundary by rejecting the sharing of ride-sharing addresses with a partner, despite allowing it for almost every other domestic contact."}

{\bf Views toward Institutions}. 
High predictability is associated with binary views toward corporations and government (e.g., high trust or total skepticism). Low-accuracy narratives tend to focus on ``neutrality," which is insufficient for prediction. \\
\underline{High accuracy:} \textit{“she prioritizes corporate transparency but has low trust in social media.”}\\
\underline{Low accuracy:} \textit{“her trust in banking institutions is neutral. She is neither deeply bothered by data collection nor entirely unconcerned.”}

{\bf Privacy Behaviors and Tech Savviness}. Predictable individuals are typically either highly tech-savvy (using VPNs and opting out) or entirely un-savvy (sharing everything). Unpredictable people apply controls sporadically or lack a clear understanding of the consequences of their settings. \\
\underline{High accuracy:} texts include mentions of \textit{“declining cookies,”} \textit{“using non-tracking browsers,”} etc. Other examples include users who \textit{“reject smart home devices entirely”}. Some users \textit{“deliberately avoid certain platforms (social media) or devices (security cameras).”} \\
\underline{Low accuracy:} \textit{“this persona believes that using fake names is better than using a VPN to achieve anonymity."} This example in which an individual employs less powerful techniques to protect their privacy might make it harder for the LLM to accurately predict their future privacy decisions.

{\bf Resignation}. This theme captures users who default to the path of least resistance for two distinct reasons: a perceived lack of control or “privacy fatigue"~\cite{privacy_fatigue}. Predictable personas handle this resignation consistently -- either by uniformly allowing / denying requests due to feeling powerless, or by always taking the easiest option due to fatigue. Unpredictable personas, however, exhibit erratic boundaries, oscillating between resignation and effortful privacy protection.
\underline{High accuracy:} \textit{``the person has zero agency and does not take complex proactive steps''} or \textit{``he feels overwhelmed by the volume of passwords he manages, leading him to write them down''}. Narratives such as these provide a stable heuristic for a model to predict privacy decisions. \\
\underline{Low accuracy:} \textit{“the individual suffers from privacy fatigue but sometimes they refuse to install an app or use an online service at all, due to concerns about privacy”} or \textit{“he feels they have no control over who accesses their location information but believes they do have some control over their private conversations (emails and messages).”}

In summary, we find that predictability fails when personas are inconsistent, misunderstand consequences, or are described in ``neutral" terms. These findings explain the individual-level performance reported in Section~\ref{sec:evaluation}. For example, Fig.~\ref{fig:split_bb_metrics_plot_aggregate} shows that the CAuthN dataset, as compared to the SPA dataset, exhibits a large number of instances where individuals chose neutral answers or had inconsistent answers about sharing data from different sensors. Similarly, we observe Pew W127 personas are less predictable on average than other datasets because of the lack of decisiveness in the responses, characterized by neutral, low-confidence, and mixed answers. 

Appendix~\ref{app:sample-personas} details two participants from the Pew PP1 survey that illustrate this contrast in predictability. Participant A, yielding high prediction accuracy (83\%), maintains a consistent, tiered approach to privacy with clear boundaries between sensitive data (e.g., SSN, health records) and less critical data—a structured posture the persona easily formalizes. In contrast, Participant B proves highly challenging to predict (50\% accuracy) due to conflicting decisions, such as avoiding mainstream social media for security while frequently ``egosurfing" and oscillating between anonymity and real names.

%% file: content/60_discussion.tex
\section{Discussion}

We discuss the broader implications of our findings for simulating user studies and enabling personalized privacy assistants, alongside the limitations of this work.

\subsubsection*{User Study Simulation}
As shown in Section~\ref{sec:cross_study}, \name can extract highly predictive personas from one study to accurately simulate population-level responses in an independent study. This capability enables the creation of \textit{synthetic panels}—collections of validated, text-based personas that represent specific target populations. Researchers and practitioners could use these panels to pilot new survey instruments, test UI variants, or evaluate experimental conditions before deploying them to participants, reducing study costs and participant fatigue. However, creating reliable synthetic panels presents two challenges. First, representing specific target populations requires access to diverse, high-fidelity datasets, while available ones are skewed heavily toward WEIRD (Western, educated, industrialized, rich, and democratic) demographics. Second, privacy preferences shift as attitudes and norms evolve alongside new technologies. A synthetic panel derived from older studies will eventually fail to capture contemporary privacy behaviors, necessitating continuous  updates.

\subsubsection*{Personalized Privacy and the Cold Start Problem}
Beyond population-level simulations, text-based personas can power future personalized privacy assistants. An assistant could utilize an individual's validated persona as a system instruction to personalize privacy decisions. Standardizing these preferences into a concise text format addresses a major hurdle for AI assistants: the \textit{cold start problem}. Currently, onboarding a user to a new privacy assistant requires either tedious  questionnaires—which often fail to predict actual behavior—or collecting massive amounts of raw behavioral data over time. In contrast, text-based personas offer inherent \textit{portability}. A persona summarized from a user's behavior on one platform can be extracted as a structured, human-readable text file and securely ported to bootstrap a new privacy assistant on an entirely different platform. This portability resolves the cold start problem without requiring the transfer of sensitive raw history logs between services.

 However, our findings point to a core challenge in designing privacy agents: perfectly replicating a user's behavior may not actually serve their best interests. Because human privacy decisions are often contradictory, poorly informed, or driven by privacy resignation, future research must move beyond merely maximizing the predictability of possibly flawed behavior. Instead, we must find ways to capture and act upon users' true preferences. Agents must distinguish between intentional choices and patterns arising from resignation or confusion. When a persona yields low-confidence predictions due to conflicting signals, agents should proactively engage the user. Without overburdening the user with frequent interruptions, this engagement can highlight personal inconsistencies and ask targeted questions to help users clarify their actual intent.

\subsubsection*{Limitations}
The evaluation of \name is subject to several limitations. First, the datasets exclusively cover adult populations in the US and UK. Expanding the geographical and cultural diversity of the validation datasets is necessary to fully assess representation. Second, the generated text schemas capture a temporal snapshot of a user's attitudes; more work is needed to develop mechanisms that continuously calibrate and update text personas over time as user preferences drift. Finally, because running the iterative optimization across five large datasets incurs prohibitive API costs (our evaluation consumed tens of billions of tokens), our results are derived primarily from the Gemini model family (Gemini 3.0 Flash and Gemini 2.5 Flash-Lite). However, concurrent work indicates that different frontier models perform similarly when simulating privacy behaviors~\cite{flemings2025, li2026llmagentssimulateenduser}.

%% file: content/70_conclusion.tex
\section{Conclusion}

Simulating individual privacy decisions requires balancing competing demands: predictiveness, auditability, scalability, and representation. While prior approaches often compromise one dimension for another, \name transforms raw survey data into concise, grounded privacy personas that balance these dimensions. By compressing inputs into structured narratives, it reduces prompt tokens by 80--95\% across five diverse datasets while preserving high predictive accuracy. Our evaluation reveals that past behaviors are stronger predictors of future choices than general privacy attitudes, and that dynamically tailoring persona templates further maximizes performance. Moreover, the resulting personas were highly transferable, remaining predictive across diverse populations and topics. Although these personas must continue to evolve alongside technological shifts, we show that scalable, accurate, and human-interpretable simulation is achievable, laying a practical foundation for transparent personalized privacy assistants.

%% file: content/71_ethics.tex
\section*{Ethics Considerations}

We use publicly available research datasets strictly in line with their original intent: understanding human privacy reasoning. \name synthesizes text-based personas using only the explicitly provided responses within each dataset. We do not attempt to cross-reference this data with external sources, de-anonymize participants, or infer information beyond the scope of their stated responses.

%% file: content/81_appendix.tex
\subsection{Detailed Results}
\label{appendix:detailed-results}

Below we report detailed results for the experiments described in Section~\ref{sec:experiments}.

\subsubsection{Number of Questions Used for Generation and Evaluation}
\label{appendix:detailed-results-question-count}

Table~\ref{tab:question_count_all} reports the average and standard deviation number of questions used for both persona generation and evaluation, across all the users in each of the datasets. In addition, Tables~\ref{tab:question_count_ab} and~\ref{tab:question_count_bb} report these statistics when generation is performed using only attitude and behavioral questions, respectively.

\begin{table}[h]
\centering
\begin{tabular}{l|cc}
\hline
Dataset & Generation & Evaluation \\
\hline
SPA & 129.1 ($\pm$15.6) & 32.8 ($\pm$3.9) \\
Pew PP1 & 40.6 ($\pm$2.8) & 10.8 ($\pm$0.8) \\
Pew W49 & 58.8 ($\pm$4.4) & 15.2 ($\pm$1.2) \\
Pew 127 & 105.0 ($\pm$6.5) & 26.8 ($\pm$1.7) \\
CAuthN & 66.0 ($\pm$0.0) & 17.0 ($\pm$0.0) \\
\hline
\end{tabular}
\caption{The mean (and standard deviation) number of questions used during generation and evaluation, across all individuals in the studies, when generating and evaluating personas using all questions in the surveys.}
\label{tab:question_count_all}
\end{table}

\begin{table}[h]
\centering
\begin{tabular}{l|cc}
\hline
Dataset & Generation & Evaluation \\
\hline
SPA & 16.0 ($\pm$0.0) & 145.9 ($\pm$19.5) \\
Pew PP1 & 37.0 ($\pm$2.2) & 6.5 ($\pm$1.1) \\
Pew W49 & 44.7 ($\pm$4.6) & 14.6 ($\pm$0.9) \\
Pew 127 & 81.1 ($\pm$6.0) & 18.0 ($\pm$2.3) \\
CAuthN & 62.0 ($\pm$0.0) & 15.0 ($\pm$0.0) \\
\hline
\end{tabular}
\caption{The mean (and standard deviation) number of questions used during generation and evaluation, across all individuals in the surveys, when generating from attitude questions and evaluating on behavioral questions.}
\label{tab:question_count_ab}
\end{table}

\begin{table}[h]
\centering
\begin{tabular}{l|cc}
\hline
Dataset & Generation & Evaluation \\
\hline
SPA & 116.3 ($\pm$15.6) & 29.6 ($\pm$3.9) \\
Pew PP1 & 4.5 ($\pm$1.1) & 2.0 ($\pm$0.1) \\
Pew W49 & 11.5 ($\pm$0.8) & 3.1 ($\pm$0.3) \\
Pew 127 & 14.1 ($\pm$1.9) & 3.9 ($\pm$0.4) \\
CAuthN & 12.0 ($\pm$0.0) & 3.0 ($\pm$0.0) \\
\hline
\end{tabular}
\caption{The mean (and standard deviation) number of questions used during generation and evaluation, across all individuals in the surveys, when generating from behavioral questions and evaluating on behavioral questions.}
\label{tab:question_count_bb}
\end{table}

\subsubsection{Impact of Optimizations Steps}
\label{appendix:detailed-results-iteration-steps}
\begin{figure}[h]
\centering
\includegraphics[width=\linewidth,keepaspectratio]{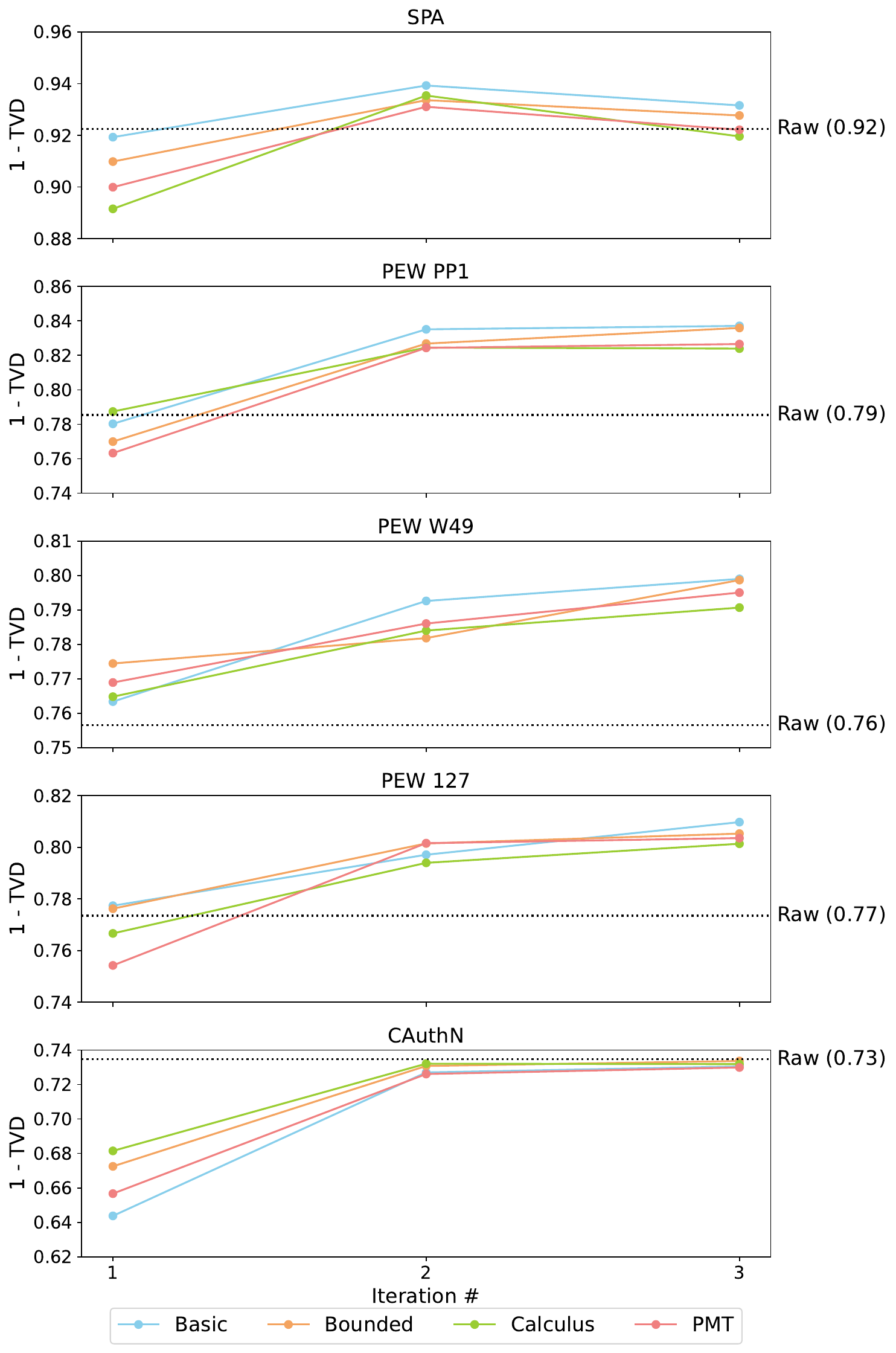}
\caption{The performance of the narratives at different iteration steps, across datasets. The plot presents the population-level metric $1-TVD$ ($\uparrow$) on the evaluation set as the number of iterations $I$ increases. We observe that the performance plateaus after two iterations, suggesting that additional optimization steps are unlikely to result in better-performing personas.}
\label{fig:tvd_iterations}
\end{figure}

We assess how the number of iterations employed in the refinement of the persona impacts their predictive power of these personas. Fig.~\ref{fig:tvd_iterations} presents the results across datasets. The figure shows that there are limited gains beyond the second iteration. This finding indicates that optimization is needed to improve the persona's performance and that \name is efficient, not requiring more than two iterations to converge.

\subsubsection{Additional Population-Level Metrics}
\label{app:more_metrics}
We adopt two additional measures from prior literature to compare the distribution of predicted responses to that of true responses:  mean estimation error from Krsteski et al.~\cite{krsteski2025} and the Wasserstein Distance (WD)~\cite{suh2025}. 

\paragraph{Mean Estimation Error}
The mean estimation error (MEE) is given as the relative error between mean estimated over the real responses and that over the predicted responses for a given question over all the individuals. Specifically, the true mean for question $q_i$ is $\mu_i = \frac{1}{M} \sum_{j=1}^{M} f_i(r_{j,i})$ and the predicted mean is $\hat{\mu}_i = \frac{1}{M} \sum_{j=1}^{M} f_i(\hat{r}_{j,i}).$
Then, $MEE_i$ for question $q_i$ is:  

$$MEE_i = 100 \times \frac{|\hat{\mu}_i - \mu_i|}{\mu_i}.$$

We can also report the macro average of the mean estimation error over all the questions in the dataset as: $$\text{MEE}_{S} = \frac{1}{|\mathcal{Q}_{eval}|} \sum_{q_i \in \mathcal{Q}_{eval}} MEE_i.$$ 

\paragraph{Wasserstein Distance}
For a given question $q_i$ and response $a$, we obtain the probability mass function of true responses as $P_i(a) = \frac{1}{M} \sum_{j=1}^{M} \mathbb{I}(f_i(r_{j,i}) = a)$ and that of predicted responses as $\hat{P}_i(a) = \frac{1}{M} \sum_{j=1}^{M} \mathbb{I}(f_i(\hat{r}_{j,i}) = a)$. Then, we estimate the cumulative distribution functions as $C_i(v) = \sum_{a=1}^{v} P_i(a)$ and $\hat{C}_i(v) = \sum_{a=1}^{v} \hat{P}_i(a)$. The Wasserstein distance $WD_i$ for question $q_i$ is then:$$WD_i = \sum_{v=1}^{m_i} | C_i(v) - \hat{C}_i(v) |.$$

Similar to the metrics, we report the macro average of the Wasserstein distance over all evaluation questions as:$$\text{WD}_{S} = \frac{1}{|\mathcal{Q}_{eval}|} \sum_{q_i \in \mathcal{Q}_{eval}} WD_i.$$

\paragraph{Weighted Total Variation Distance}
We extend the Total Variation Distance (TVD) computation by incorporating respondent weights. In the cross-study setting, we compare the simulated responses generated by personas derived from a source study against the real responses from a target study. For a given question $q_i$ in the target study and response value $a$, let $w_j^{(T)}$ denote the survey weight assigned to respondent $j$ in the target study. We compute the weighted probability mass function of true responses as:

$$P_i(a) = \frac{\sum_{j=1}^{M_T} w_j^{(T)} \mathbb{I}(f_i(r_{j,i}) = a)}{\sum_{j=1}^{M_T} w_j^{(T)}}$$

where $M_T$ is the total number of respondents in the target study, and $r_{j,i}$ is the true response of target respondent $j$ to question $q_i$.

Similarly, let $w_k^{(S)}$ denote the survey weight assigned to the source respondent from whom persona $k$ was derived. The weighted probability mass function of predicted responses is computed as:

$$\hat{P}_i(a) = \frac{\sum_{k=1}^{M_S} w_k^{(S)} \mathbb{I}(f_i(\hat{r}_{k,i}) = a)}{\sum_{k=1}^{M_S} w_k^{(S)}},$$

where $M_S$ is the total number of personas (derived from the source study), and $\hat{r}_{k,i}$ is the predicted response of persona $k$ to question $q_i$.

The Total Variation Distance $\text{TVD}_i$ for question $q_i$ is then calculated using these two distinct weighted distributions:

$$\text{TVD}_i = \frac{1}{2} \sum_{a=1}^{m_i} | P_i(a) - \hat{P}_i(a) |$$

where $m_i$ is the number of possible answers for question $q_i$. The aggregated metric $\text{TVD}_{S}$ and the average TVComplement are then computed across all evaluation questions $\mathcal{Q}_{eval}$ as described in the main text.

\subsubsection{Influence of Privacy Behavior Theories}

\begin{figure}[h]
\centering
\includegraphics[width=\linewidth,keepaspectratio]{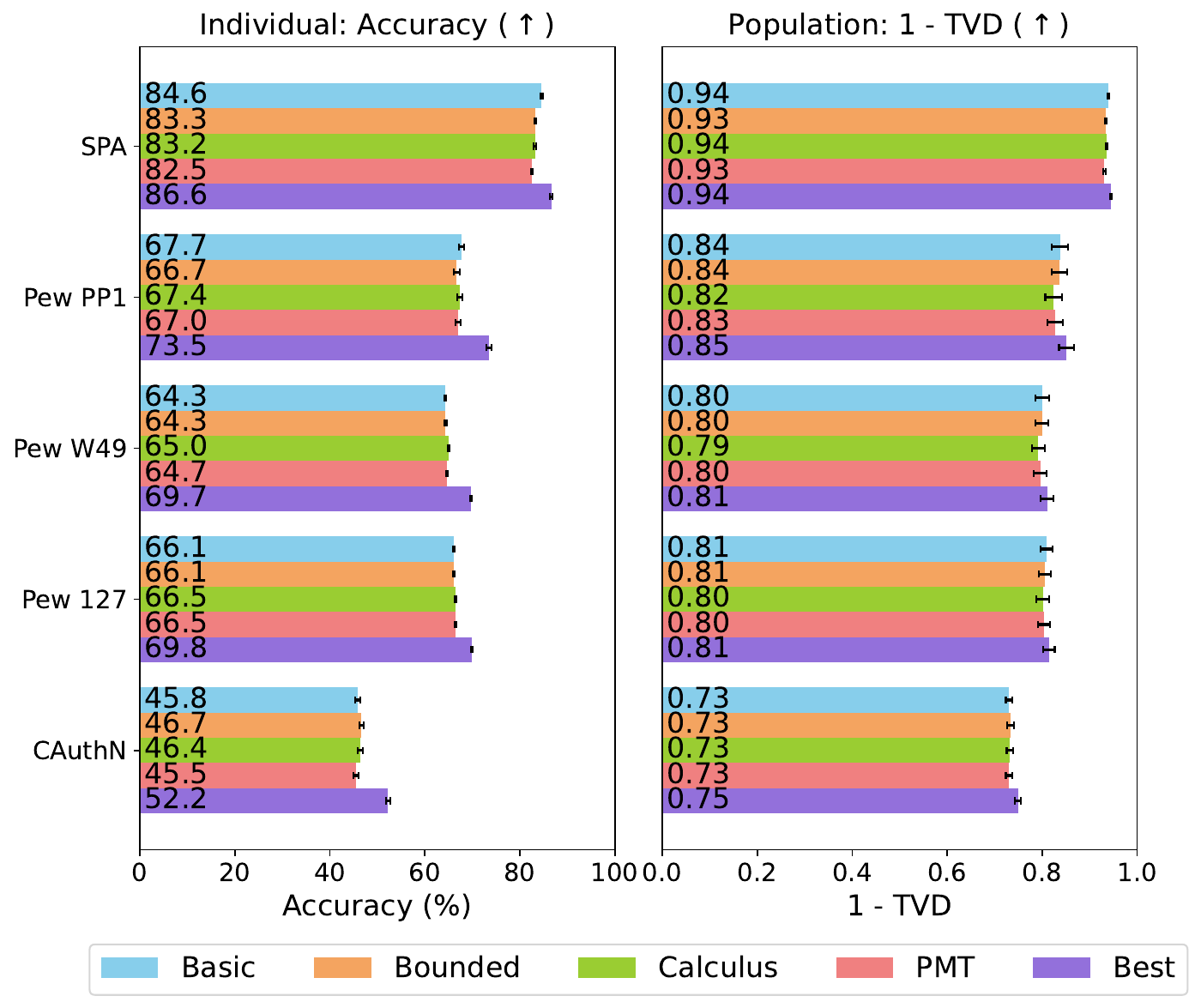}
\caption{Individual- and population level performance across prompt templates inspired by privacy behavior theories. Results using 80:20 generation/evaluation split. Error bars represent 95\% confidence intervals obtained through bootstrapping the results 1,000 times with replacement.}
\label{fig:split_0.2_metrics_plot}
\end{figure}

\begin{table}[t]
    \centering
    \resizebox{\columnwidth}{!}{%
\begin{tabular}{  c  c c c c}%
\toprule
 \textbf{Dataset} & \textbf{Base} & \textbf{Bounded} & \textbf{Calculus} & \textbf{PMT} \\ 
 \midrule
SPA & 26.59 ($\pm$1.88) & 29.15 ($\pm$1.77) & 30.10 ($\pm$1.99) & 31.47 ($\pm$1.80) \\
Pew PP1 & 21.55 ($\pm$7.63) & 21.69 ($\pm$6.56) & 25.81 ($\pm$8.89) & 22.14 ($\pm$6.66) \\
Pew W49 & 25.98 ($\pm$3.85) & 26.22 ($\pm$3.92) & 27.48 ($\pm$4.06) & 26.22 ($\pm$3.74) \\
Pew W127 & 18.72 ($\pm$2.60) & 18.56 ($\pm$2.50) & 18.84 ($\pm$2.63) & 19.00 ($\pm$2.64) \\
CAuthN & 15.69 ($\pm$0.63) & 15.40 ($\pm$0.68) & 16.23 ($\pm$0.68) & 17.31 ($\pm$0.77) \\
 \bottomrule
\end{tabular}
}
\caption{Impact of persona-generating prompt on the population-level metric Mean Estimation Error $MEE_S$ ($\downarrow$). Each cell also reports 95\% confidence intervals obtained through bootstrapping the results 1,000 times with replacement. The performance is similar across the datasets, suggesting that the high predictiveness of personas does not stem, on average, from the choice in the prompt.}
\label{tab:narrative_template_choice_performance_mee}
\end{table}

\begin{table}[h]
    \centering
    \setlength{\tabcolsep}{3.5pt} %
\begin{tabular}{ c c c c c }%
\toprule
 \textbf{Dataset} & \textbf{Base} & \textbf{Bounded} & \textbf{Calculus} & \textbf{PMT} \\ 
 \midrule
SPA & 0.06 ($\pm$0.00) & 0.07 ($\pm$0.00) & 0.07 ($\pm$0.00) & 0.07 ($\pm$0.00) \\
Pew PP1 & 0.21 ($\pm$0.02) & 0.21 ($\pm$0.02) & 0.22 ($\pm$0.02) & 0.22 ($\pm$0.02) \\
Pew W49 & 0.22 ($\pm$0.02) & 0.24 ($\pm$0.02) & 0.24 ($\pm$0.01) & 0.23 ($\pm$0.02) \\
Pew 127 & 0.24 ($\pm$0.02) & 0.24 ($\pm$0.02) & 0.25 ($\pm$0.02) & 0.25 ($\pm$0.02) \\
CAuthN & 0.65 ($\pm$0.03) & 0.62 ($\pm$0.03) & 0.63 ($\pm$0.03) & 0.66 ($\pm$0.03) \\
 \bottomrule
\end{tabular}
\caption{Impact of the persona-generating prompt on the population-level metric $WD_S$ ($\downarrow$), on the five datasets. The performance is similar across prompt variants for each datasets, suggesting that the high predictiveness of the personas does not stem from the choice in the prompt.}
\label{tab:narrative_template_choice_performance_wd}
\end{table}

Fig.~\ref{fig:split_0.2_metrics_plot} highlights the detailed individual-level and population-level ($1-TVD$) performance across prompts from the four privacy theories: Privacy Calculus, Bounded Rationality and Protection Motivation Theory (PMT). The prompts we create for these theories are listed in Appendix~\ref{appendix:prompts}. The remaining population-level metrics ($1-TVD$, $MEE_S$, and $WD_S$) are reported in Tables~\ref{tab:narrative_template_choice_performance_tvd},~\ref{tab:narrative_template_choice_performance_mee}, and ~\ref{tab:narrative_template_choice_performance_wd}, respectively.

\subsubsection{Influence of Attitude and Behavioral Questions}

\paragraph{Generating from Attitude Questions and Evaluating on Behavioral Questions}

\begin{figure}[h]
\centering
\includegraphics[width=\linewidth,keepaspectratio]{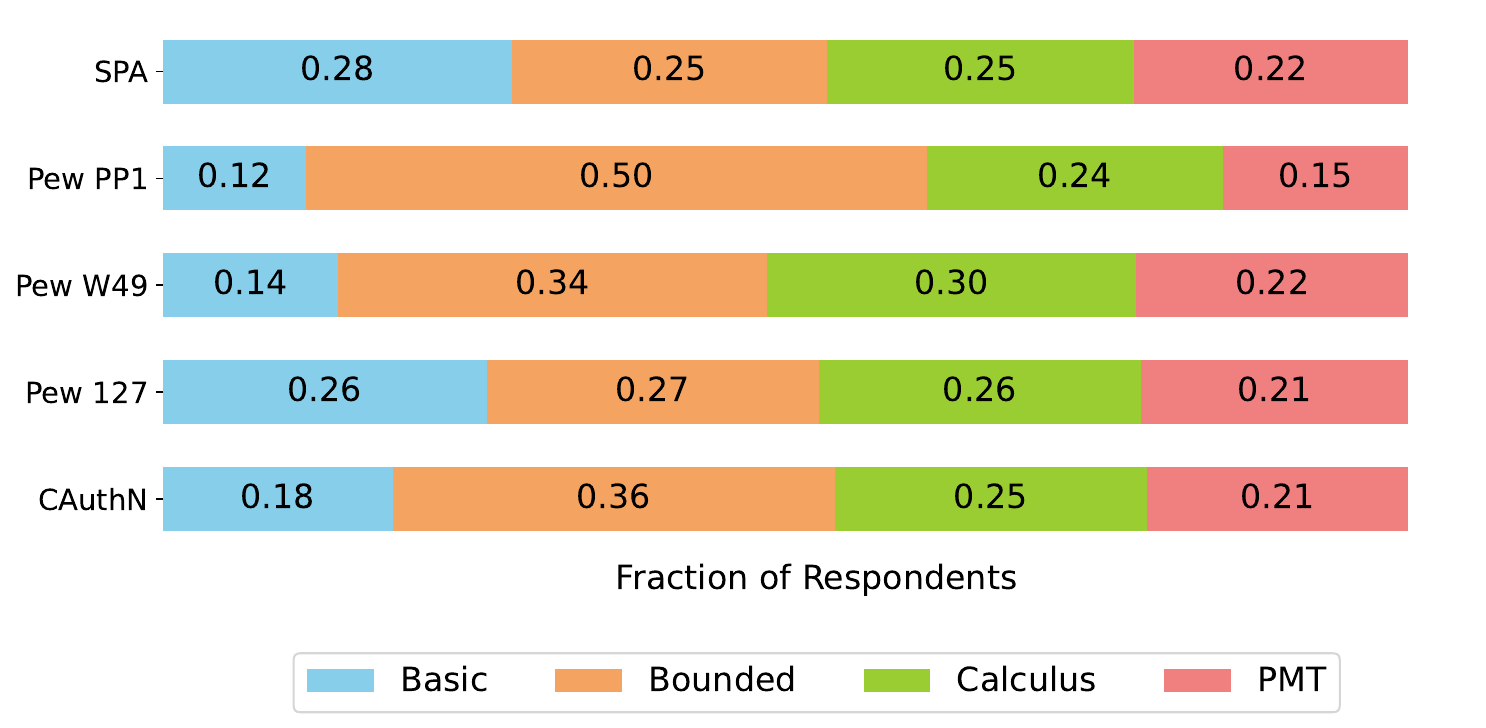}
\caption{The fraction of individuals for which each of the templates is best performing when generating from attitude questions and evaluating on behavioral questions.}
\label{fig:split_ab_metrics_best_synthesizer}
\end{figure}

Fig.~\ref{fig:split_ab_metrics_best_synthesizer} illustrates the fraction of individuals for whom each specific template yielded the highest predictive accuracy when the generation set consists of attitude questions and the evaluation set contains only behavioral questions.

\begin{figure}[h]
\centering
\includegraphics[width=\linewidth,keepaspectratio]{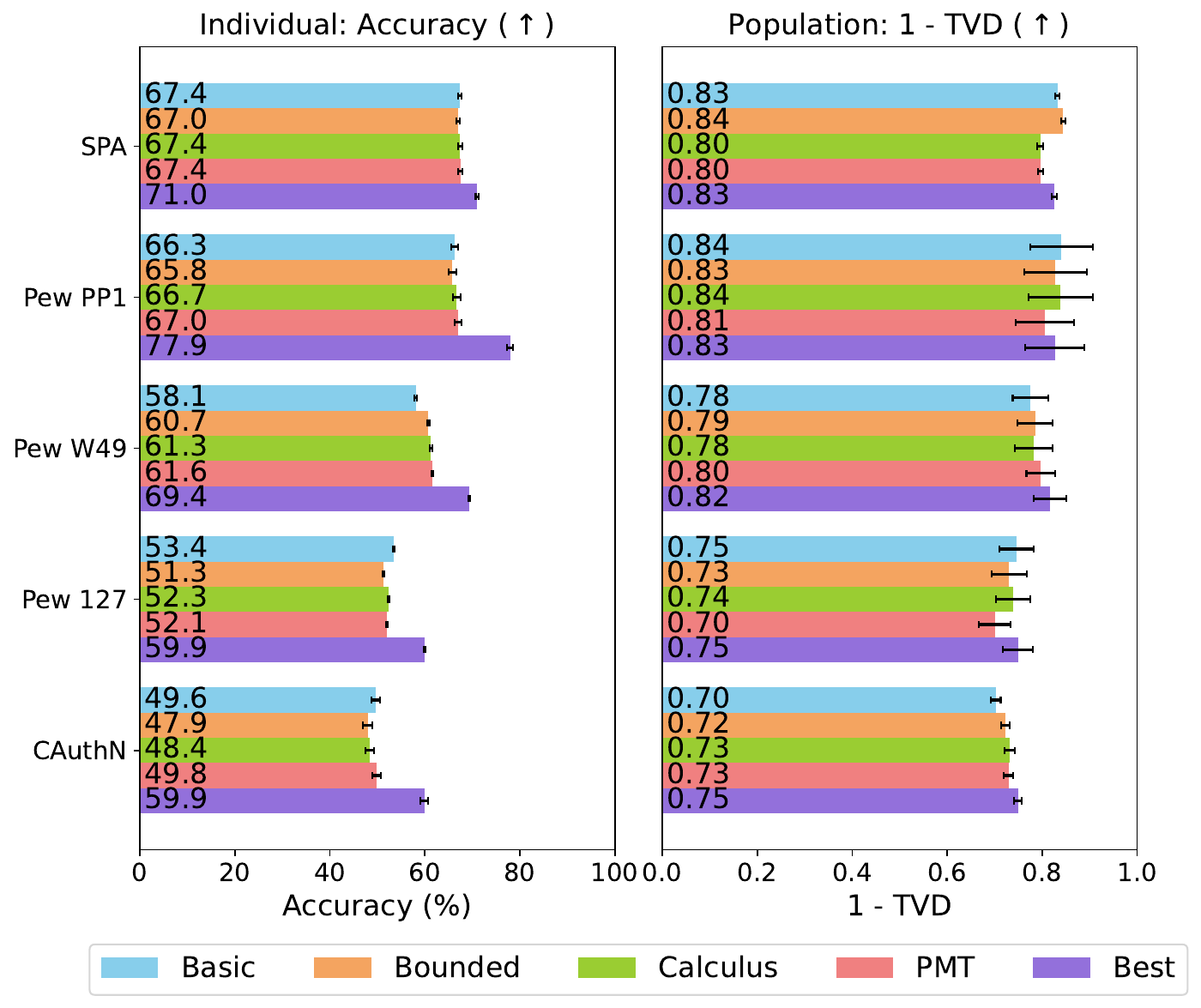}
\caption{Individual- and population level performance across prompt templates inspired by privacy behavior theories. Results when generating from attitude questions and evaluating on behavioral questions. Error bars represent 95\% confidence intervals obtained through bootstrapping the results 1,000 times with replacement.}
\label{fig:split_ab_metrics_plot}
\end{figure}

\begin{table}[h]
    \centering
    \setlength{\tabcolsep}{3.5pt} %
\begin{tabular}{  c  c c c c}%
\toprule
 \textbf{Dataset} & \textbf{Base} & \textbf{Bounded} & \textbf{Calculus} & \textbf{PMT} \\ 
 \midrule
SPA & 0.17 ($\pm$0.00) & 0.16 ($\pm$0.00) & 0.20 ($\pm$0.01) & 0.20 ($\pm$0.01) \\
Pew PP1 & 0.16 ($\pm$0.07) & 0.17 ($\pm$0.06) & 0.16 ($\pm$0.07) & 0.19 ($\pm$0.07) \\
Pew W49 & 0.22 ($\pm$0.04) & 0.21 ($\pm$0.03) & 0.22 ($\pm$0.04) & 0.20 ($\pm$0.03) \\
Pew 127 & 0.25 ($\pm$0.04) & 0.27 ($\pm$0.04) & 0.26 ($\pm$0.04) & 0.30 ($\pm$0.03) \\
CAuthN & 0.30 ($\pm$0.01) & 0.28 ($\pm$0.01) & 0.27 ($\pm$0.01) & 0.27 ($\pm$0.01) \\
 \bottomrule
\end{tabular}
\caption{Impact of persona-generating prompt on the population-level metric $TVD$ ($\downarrow$), on the five datasets. Each cell also reports 95\% confidence intervals obtained through bootstrapping the results 1,000 times with replacement. Results are reported on the five datasets when generating from attitude questions and evaluating on behavioral questions.}
\label{tab:narrative_template_choice_performance_tvd_ab}
\end{table}

\begin{table}[h]
    \centering
        \resizebox{\columnwidth}{!}{%
\begin{tabular}{c c c c c}%
\toprule
 \textbf{Dataset} & \textbf{Base} & \textbf{Bounded} & \textbf{Calculus} & \textbf{PMT} \\ 
 \midrule
SPA & 51.01 ($\pm$1.27) & 50.31 ($\pm$1.36) & 58.34 ($\pm$0.95) & 56.49 ($\pm$0.97) \\
Pew PP1 & 38.85 ($\pm$12.78) & 44.07 ($\pm$13.12) & 32.32 ($\pm$13.31) & 55.39 ($\pm$13.88) \\
Pew W49 & 44.81 ($\pm$9.77) & 40.16 ($\pm$7.02) & 48.97 ($\pm$12.58) & 39.80 ($\pm$6.56) \\
Pew 127 & 32.62 ($\pm$5.80) & 34.03 ($\pm$6.78) & 35.22 ($\pm$7.64) & 39.73 ($\pm$7.86) \\
CAuthN & 25.89 ($\pm$1.13) & 25.64 ($\pm$1.35) & 25.03 ($\pm$1.25) & 23.98 ($\pm$1.05) \\
 \bottomrule
\end{tabular}
}
\caption{Impact of persona-generating prompt on the population-level metric $MEE_S$ ($\downarrow$), on the five datasets. Each cell also reports 95\% confidence intervals obtained through bootstrapping the results 1,000 times with replacement. Results are reported on the five datasets when generating from attitude questions and evaluating on behavioral questions.}
\label{tab:narrative_template_choice_performance_mee_ab}
\end{table}

\begin{table}[h]
    \centering
\begin{tabular}{ c c c c c }%
\toprule
 \textbf{Dataset} & \textbf{Base} & \textbf{Bounded} & \textbf{Calculus} & \textbf{PMT} \\ 
 \midrule
SPA & 0.17 ($\pm$0.00) & 0.16 ($\pm$0.00) & 0.20 ($\pm$0.01) & 0.20 ($\pm$0.01) \\
Pew PP1 & 0.17 ($\pm$0.08) & 0.18 ($\pm$0.08) & 0.17 ($\pm$0.08) & 0.20 ($\pm$0.07) \\
Pew W49 & 0.26 ($\pm$0.06) & 0.24 ($\pm$0.05) & 0.23 ($\pm$0.05) & 0.22 ($\pm$0.04) \\
Pew 127 & 0.33 ($\pm$0.07) & 0.35 ($\pm$0.06) & 0.31 ($\pm$0.07) & 0.36 ($\pm$0.06) \\
CAuthN & 0.72 ($\pm$0.03) & 0.68 ($\pm$0.03) & 0.68 ($\pm$0.04) & 0.65 ($\pm$0.03) \\
 \bottomrule
\end{tabular}
\caption{Impact of persona-generating prompt on the population-level metric $WD_S$ ($\downarrow$), on the five datasets. Each cell also reports 95\% confidence intervals obtained through bootstrapping the results 1,000 times with replacement. Results are reported on the five datasets when generating from attitude questions and evaluating on behavioral questions.}
\label{tab:narrative_template_choice_performance_wd_ab}
\end{table}

For the setting in which only attitude questions are used for generation, Fig.~\ref{fig:split_ab_metrics_plot} highlights the detailed individual-level and population-level ($1-TVD$) performance across prompts from the four privacy theories: Privacy Calculus, Bounded Rationality and Protection Motivation Theory (PMT). The prompts we create for these theories are listed in Appendix~\ref{appendix:prompts}. The population-level metrics ($TVD$, $MEE_S$, and $WD$) are reported in Tables,~\ref{tab:narrative_template_choice_performance_tvd_ab},~\ref{tab:narrative_template_choice_performance_mee_ab}, and~\ref{tab:narrative_template_choice_performance_wd_ab},  respectively.

\paragraph{Generating from Behavioral Questions and Evaluating on Behavioral Questions}

\begin{figure}[ht]
\centering
\includegraphics[width=\linewidth,keepaspectratio]{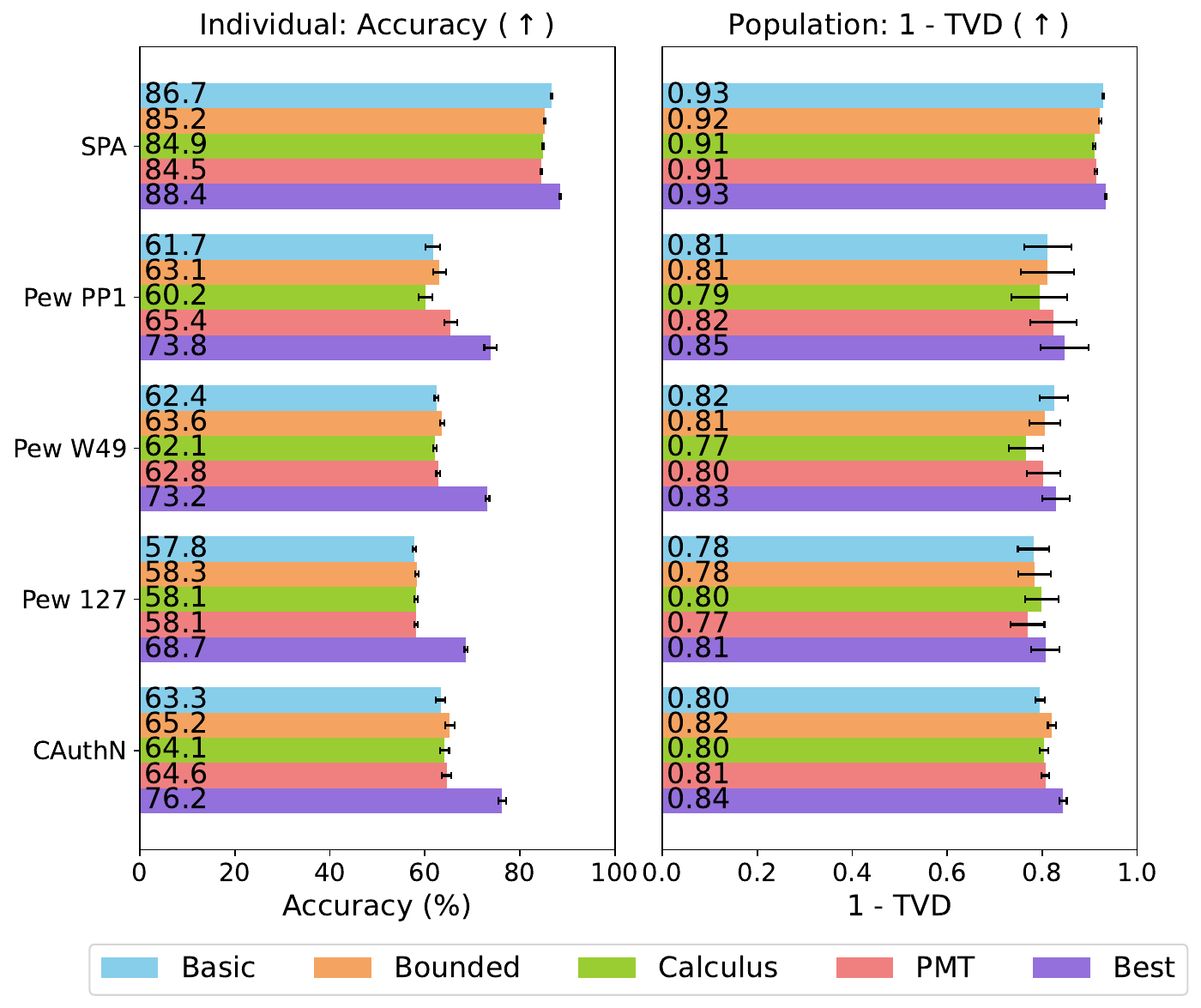}
\caption{Individual- and population level performance across prompt templates inspired by privacy behavior theories. Results when generating from behavioral questions and evaluating on behavioral questions. Error bars represent 95\% confidence intervals obtained through bootstrapping the results 1,000 times with replacement.}
\label{fig:split_bb_metrics_plot}
\end{figure}

For the setting in which only behavioral questions are used for generation and evaluation, Fig.~\ref{fig:split_bb_metrics_plot} highlights the detailed individual-level and population-level ($1-TVD$) performance across prompts from the four privacy theories: Privacy Calculus, Bounded Rationality and Protection Motivation Theory (PMT). The prompts we create for these theories are listed in Appendix~\ref{appendix:prompts}. The remaining population-level metrics ($MEE_S$ and $TVD$) are reported in Tables~\ref{tab:narrative_template_choice_performance_mee_bb} and ~\ref{tab:narrative_template_choice_performance_tvd_bb}, respectively.

\begin{table}[h]
    \centering
  \resizebox{\columnwidth}{!}{%
\begin{tabular}{  c  c c c c}%
\toprule
 \textbf{Dataset} & \textbf{Base} & \textbf{Bounded} & \textbf{Calculus} & \textbf{PMT} \\ 
 \midrule
SPA & 31.63 ($\pm$1.71) & 34.52 ($\pm$2.00) & 36.89 ($\pm$1.56) & 37.51 ($\pm$1.76) \\
Pew PP1 & 77.42 ($\pm$37.57) & 66.72 ($\pm$27.41) & 101.97 ($\pm$57.18) & 50.42 ($\pm$15.60) \\
Pew W49 & 39.35 ($\pm$8.72) & 36.72 ($\pm$7.76) & 53.54 ($\pm$9.96) & 42.47 ($\pm$7.49) \\
Pew 127 & 27.44 ($\pm$3.52) & 28.81 ($\pm$4.11) & 23.91 ($\pm$3.48) & 30.65 ($\pm$4.28) \\
CAuthN & 16.80 ($\pm$1.07) & 12.39 ($\pm$0.90) & 14.51 ($\pm$0.97) & 13.66 ($\pm$0.99) \\
 \bottomrule
\end{tabular}
}
\caption{Impact of persona-generating prompt on the population-level metric $MEE_S$ ($\downarrow$), on the five datasets. Each cell also reports 95\% confidence intervals obtained through bootstrapping the results 1,000 times with replacement. Results are reported on the five datasets when generating from behavioral questions and evaluating on behavioral questions.}
\label{tab:narrative_template_choice_performance_mee_bb}
\end{table}

\begin{table}[h]
    \centering
    \setlength{\tabcolsep}{3.5pt} %
\begin{tabular}{  c  c c c c}%
\toprule
 \textbf{Dataset} & \textbf{Base} & \textbf{Bounded} & \textbf{Calculus} & \textbf{PMT} \\ 
 \midrule
SPA & 0.07 ($\pm$0.00) & 0.08 ($\pm$0.00) & 0.09 ($\pm$0.00) & 0.09 ($\pm$0.00) \\
Pew PP1 & 0.19 ($\pm$0.05) & 0.19 ($\pm$0.06) & 0.21 ($\pm$0.06) & 0.18 ($\pm$0.05) \\
Pew W49 & 0.18 ($\pm$0.03) & 0.19 ($\pm$0.03) & 0.23 ($\pm$0.04) & 0.20 ($\pm$0.04) \\
Pew 127 & 0.22 ($\pm$0.03) & 0.22 ($\pm$0.03) & 0.20 ($\pm$0.04) & 0.23 ($\pm$0.04) \\
CAuthN & 0.20 ($\pm$0.01) & 0.18 ($\pm$0.01) & 0.20 ($\pm$0.01) & 0.19 ($\pm$0.01) \\
 \bottomrule
\end{tabular}
\caption{Impact of persona-generating prompt on the population-level metric $TVD$ ($\downarrow$), on the five datasets. Each cell also reports 95\% confidence intervals obtained through bootstrapping the results 1,000 times with replacement. Results are reported on the five datasets when generating from behavioral questions and evaluating on behavioral questions.}
\label{tab:narrative_template_choice_performance_tvd_bb}
\end{table}

\subsection{Privacy Theories}
\label{appendix:privacy_theories}
\paragraph{Base Persona}
The first template is a basic one that serves as a baseline persona representation. It has two sets of instructions. First, it instructs the LLM to analyze the behavior of an individual based on their responses ($\textbf{r}_{gen}$). It then instructs the LLM to output a concise and precise persona narrative that explains the logic behind the individual's responses. The instructions utilize chain-of-thought prompting to generate the persona, check if it answers the given responses correctly, and update it to correctly answer the questions.

\paragraph{Privacy Calculus}
Research around how individuals disclose private information has theorized that individuals follow a privacy calculus~\cite{privacy_calculus}. Laufer and Wolfe studied present privacy as an information management process and coined the term ``calculus of behavior,"~\cite{calculus_1977} where individuals assess the consequences of information disclosure in terms of future risks and benefits. In particular, individuals can disclose what they consider to be private information if they perceive benefits. Later, Culnan and Armstrong presented privacy calculus as a process where individuals rationally decide to disclose information if they perceive benefits as exceeding the risks. Knijnenburg et al. proposed to operationalize privacy calculus for personalized privacy~\cite{knijnenburg2017death}. Inspired by their model, this template models the privacy calculus of an individual by following the process: identifying contextual variables that influence individual's privacy posture, determining risk assessment method of how does the individual determine risk, determining their benefit drivers of what types of benefits are most compelling to them, and extracting their decision strategy of how they compare risks to benefits. 

\paragraph{Bounded Rationality}
While privacy calculus assumes that individuals are rational actors in privacy decision making, other research shows that such privacy decisions are limited by individuals' bounded rationality~\cite{bounded_rational_1, bounded_rational_2}. In that framework, individuals are incapable of performing a proper risk assessment because some of the risks are invisible to them and they do not have enough knowledge to estimate risks. Besides, individuals favor immediate gratification from information disclosure over distant and often undefined privacy risks. This template models the bounded rationality of an individual by following this process: determining cognitive biases, such as optimism bias or present bias that shape individual perception of risk; identifying what triggers (e.g., excitement for a new feature) tend to override rational decision making; and describing the primary heuristic for privacy decision making.

\paragraph{Protection Motivation Theory (PMT)}
The protection motivation theory posits that individuals engage in protective behavior, such as not disclosing information, in response to threatening events~\cite{Rogers01091975}. In particular, individuals consider three components of fear corresponding to an event: magnitude of harm, occurrence, and effectiveness of a protective behavior. Then, they engage in a mediating process that assesses the severity of harm, expected exposure of the event, and the belief that the protective behavior is effective. The assessment centers around whether the protective behavior can remedy the expected harm from an event. This template operationalizes PMT in the context of persona generation through this process: appraise their risk of events in terms of severity and vulnerability (chance of happening), determine the coping appraisal of protective behavior (self-efficacy and effectiveness), and describe how the balance between their threat and coping appraisals determines their motivation to adopt the protective behavior.

\subsection{Dataset Description}
\label{sec:app:datasets}

\paragraph{SPA Dataset~\cite{spa_dataset}} 
Abdi et al. published a dataset about privacy norms for smart home personal assistants (SPAs)~\cite{spa_dataset}. This dataset comprises responses from 1,737 participants about their comfort with a set of 24 data sharing scenarios, their privacy attitudes measured using the 10-item IUIPC scale (10 questions)~\cite{IUIPC}, and their security attitudes (6 questions) measured using the SA-6 scale~\cite{sa6_scale}. The participants were recruited from the Prolific service prior to 2021. Abdi et al. do not indicate the country of the respondents and the precise time of data collection.  

The scenarios are sampled from a larger set of 120 scenarios covering whether the smart assistant should share data with skills. These scenarios, constructed to cover 15 data types, have eight scenarios for each data type. The eight scenarios correspond to providing a purpose, omitting a purpose, and varying 6 types of recipients. Each scenario has a set of questions where the respondent indicates their comfort, on a five-point Likert scale, with the personal assistant sharing a specific data type with different recipients, purposes, or conditions. 

When analyzing this dataset, we partition the responses into three groups: behaviors, privacy attitudes, and security attitudes. The first group includes the responses to the data sharing scenarios (144 separate responses on average per respondent). We use the first group to represent personalized privacy use cases, where we group the responses into two binary categories: acceptable and unacceptable. We discard the Neutral answers from the analysis. The two other groups (16 responses) denote the responses to the privacy and security attitude questionnaires. 

\paragraph{Pew PP1~\cite{pew_wpp1}} %
The Pew Research Center conducted this study between January 10-27, 2014, as part of the first American Trends Panel (ATP). The dataset includes responses from 607 US adults regarding their perceptions of privacy, surveillance, and data management in the post-Snowden era. The study covers several dimensions of privacy, including (1) the perceived importance of controlling personal information (11 questions), (2) the level of concern regarding government and corporate monitoring (6 questions), and (3) the perceived sensitivity of various types of personal data, such as social security numbers, health information, and relationship history (10 questions).

\paragraph{Pew W49~\cite{pew_w49}} %
This dataset, titled "Americans and Privacy," was collected by Pew between June 3-17, 2019, as part of the American Trends Panel (ATP) Wave 49. It includes responses from 4,272 U.S. adults and focuses on the trade-offs inherent in the modern data ecosystem. The study is structured into sections covering (1) the perceived risks versus benefits of data collection by companies and the government (2 questions), (2) the level of concern regarding how personal data is used by various entities (multiple questions), (3) the perceived sense of control over different data types, such as physical location, search terms, and social media posts (multiple questions), and (4) the understanding of privacy policies and laws (multiple questions). The dataset consists of a series of six hypothetical scenarios where respondents were asked to rate the "acceptability" of data sharing in exchange for specific services or societal benefits (e.g., a fitness tracker sharing data with health researchers or a grocery store using loyalty card data).

\paragraph{Pew 127~\cite{pew_w127}} %
The dataset was collected by the Pew Research Center between May 15-21, 2023, as part of the American Trends Panel (ATP) Wave 127. The study includes responses from a nationally representative sample of 5,101 U.S. adults, featuring oversamples of Hispanic, Black, and Asian respondents to ensure demographic precision. The survey is structured into sections covering (1) general concerns regarding government and corporate data collection (multiple questions), (2) the perceived level of control over personal data and understanding of privacy laws (multiple questions), (3) individual digital privacy behaviors, such as the use of password managers, engagement with privacy policies, and responses to data breaches, and (4) specific attitudes toward emerging technologies, including artificial intelligence and law enforcement's use of residential cameras or private messaging data.

\paragraph{CAuthN Study dataset~\cite{cauthn_dataset}}
    Dehling et al. studied individuals' perception around continuous biometric authentication. They recruited 830 participants using the Prolific service between September and October 2022. The recruited participants are US Internet users and are balanced by sex. Each respondent is assigned a single condition about the type of a service provider. The study comprises 8 sections which cover (1) the willingness to continuously share 15 types of biometric data on a scale from 1 to 100 (15 questions), (2) the belief in the effectiveness of biometric authentication in improving security on a scale from 1 to 100 (15 questions, one per data type) efficacy, (3) perceived privacy risk from disclosing biometric data to a service provider on a scale from 1 to 100 (15 questions), (4) the perceived risk in the service provider on a 7-point Likert scale (5 questions), (5) the perceived trust in the service provider on a 7-point Likert scale (5 questions), (6) the perceived loss in unauthorized access to biometric data on a 7-point Likert scale (5 questions), and (7) the perceived susceptibility of unauthorized access to biometric data on a 5-point Likert scale (3 questions). The last section covers the privacy (8-item IUIPC scale) and security (SA-6 scale) attitude questions (14 questions) as the SPA dataset. When analyzing this dataset, we discard all responses where the response is neutral (middle point in the Likert scale).

\subsection{Prompts}
\label{appendix:prompts}

Below we provide the different prompts used for persona generation and prediction.

\subsubsection{Prompts for Persona Generation}
\label{appendix:generation_prompts}

\begin{promptbox}[prompt:basic]{Template for narrative generation: Basic}
You are analyzing the behavior of a specific user based on their history of decisions.

**User History:**
{{ raw\_narrative }}

**Task:**
Write a concise "Persona Narrative" that explains the logic behind this user's decisions.
The narrative should be precise enough to predict how this user would answer future questions.

Think silently and follow this process:
1. generate the narrative for all scenarios and user responses
2. check if the narrative predicts the answer to each scenario
3. update the narrative so that it correctly predicts the answer each scenario

***Output only the generated Narrative. Do not output thoughts or any other text (such as introductory text)***
\end{promptbox}

\begin{promptbox}[prompt:bounded]{Template for narrative generation: Bounded}
You are analyzing the behavior of a specific user based on their history of decisions.

**User History:**
{{ raw\_narrative }}

**Task:**
Write a concise "Persona Narrative" that explains the logic behind this user's decisions.

Think silently and follow this process:
1. generate the narrative for all scenarios and user responses
2. check if the narrative predicts the answer to each scenario
3. update the narrative so that it correctly predicts the answer each scenario

**Crucial Constraints:**
1.  The generated persona MUST be *fully consistent* with the answers to the survey.
2.  Follow the format below.

**Persona Narrative Format:**

This detailed privacy persona is to be based on the principles of Bounded Rationality. The persona's decisions should be predictably shaped by cognitive biases and heuristics, not a rational cost-benefit analysis. Focus on their characteristics, behaviors, and internal decision-making shortcuts.

The persona should be structured as follows:

1. Measure (User Profile \& Behaviors)

User Characteristics: Define the persona's key demographic and psychographic traits.

User Behaviors: Describe the persona's common privacy-related behaviors, especially those that appear habitual or automatic.

2. Model (The Persona's Decision-Making Heuristics)

Contextual Variables:

Choice Framing: How does the presentation of the data request (e.g., framed as a gain vs. a loss, complex vs. simple) affect their choice?

Recipient Framing: How does the perceived authority or friendliness of the recipient trigger mental shortcuts (e.g., trust heuristics)?

Risk \& Benefit Perception:

Dominant Cognitive Biases: How do biases like Optimism Bias, Present Bias (instant gratification), or the Illusion of Control shape their perception of risk?

Affective Triggers: What emotional responses (e.g., excitement for a new feature, fear of missing out) are most compelling and tend to override rational thought?

Heuristic Model:

Decision Strategy: Describe their primary heuristic (e.g., "path of least resistance," "go with the default," "trust gut feeling").

Example:

Heuristic-Driven: 'They will always click "Accept All" on cookie banners without reading because the immediate benefit of accessing the site outweighs the abstract future risk (Present Bias).'

Framing-Dependent: 'They are more likely to enable location tracking if it is framed as "gaining personalized recommendations" rather than "losing locational privacy."'

***IMPORTANT OUTPUT CONSTRAINT: ***
***Output only the generated Narrative. Do not output thoughts or any other text (such as introductory text)***
\end{promptbox}

\begin{promptbox}[prompt:calculus]{Template for narrative generation: Calculus}
You are analyzing the behavior of a specific user based on their history of decisions.

**User History:**
{{ raw\_narrative }}

**Task:**
Write a concise "Persona Narrative" that explains the logic behind this user's decisions.

Think silently and follow this process:
1. generate the narrative for all scenarios and user responses
2. check if the narrative predicts the answer to each scenario
3. update the narrative so that it correctly predicts the answer each scenario

**Crucial Constraints:**
1.  The generated persona MUST be *fully consistent* with the answers to the survey.
2.  Follow the format below.

**Persona Narrative Format:**

This detailed privacy persona is to be based on the principles of the Privacy Calculus Theory. The persona should be a complete representation of an individual's privacy decision-making process, focusing on their characteristics, behaviors, and internal 'privacy calculus.'

The persona should be structured as follows:

1. Measure (User Profile \& Behaviors)

User Characteristics: Define the persona's key demographic and psychographic traits (e.g., age, occupation, tech-savviness, general attitude towards data privacy).

User Behaviors: Describe the persona's common privacy-related behaviors and digital footprint (e.g., what social media they use, how they shop online, their app usage patterns).

2. Model (The Persona's Privacy Calculus)

Contextual Variables:

Data Sensitivity: How does their willingness to disclose information change based on the type of data being requested (e.g., email address vs. real-time location vs. biometric data)?

Recipient Trust: How does their willingness to share change based on the recipient (e.g., a friend, their bank, a new mobile game, a large tech company)?

Risk \& Benefit Determination:

Risk Assessment Method: How does the persona primarily determine risk? (e.g., based on expert opinions from news articles, their own perceptions and feelings, or the behavior of their peers).

Primary Benefit Drivers: What types of benefits are most compelling to them? (e.g., system utility and personalization, convenience, social connection, financial discounts).

Tradeoff Model:

Decision Strategy: Describe whether the persona uses a compensatory or non-compensatory decision strategy.

Example:

Compensatory: 'They will share their browsing history if it results in a significant discount on a product they want.'

Non-Compensatory: 'They will never share their contact list with an app, regardless of the features offered.'

***IMPORTANT OUTPUT CONSTRAINT: ***
***Output only the generated Narrative. Do not output thoughts or any other text (such as introductory text)***
\end{promptbox}

\begin{promptbox}[prompt:pmt]{Template for narrative generation: PMT}
You are analyzing the behavior of a specific user based on their history of decisions.

**User History:**
{{ raw\_narrative }}

**Task:**
Write a concise "Persona Narrative" that explains the logic behind this user's decisions.

Think silently and follow this process:
1. generate the narrative for all scenarios and user responses
2. check if the narrative predicts the answer to each scenario
3. update the narrative so that it correctly predicts the answer each scenario

**Crucial Constraints:**
1.  The generated persona MUST be *fully consistent* with the answers to the survey.
2.  Follow the format below.

**Persona Narrative Format:**

This detailed privacy persona is to be based on the principles of Protection Motivation Theory (PMT). The persona's privacy-protective behavior (or lack thereof) should be a function of their threat appraisal and coping appraisal. Focus on their characteristics, behaviors, and their internal motivation process.

The persona should be structured as follows:

1. Measure (User Profile \& Behaviors)

User Characteristics: Define the persona's key traits (e.g., tech-savviness, general anxiety level, exposure to privacy news).

User Behaviors: Describe their current privacy-protective behaviors (e.g., use of 2FA, password habits, software updates).

2. Model (The Persona's Threat \& Coping Appraisal)

Contextual Variables:

Threat Context: How does the type of threat (e.g., identity theft vs. social embarrassment) affect their appraisal?

Source of Information: How does the source of a threat warning (e.g., a news report, a system notification, a friend) influence their appraisal?

Appraisal Process:

Threat Appraisal: How does the persona assess threat severity ("how bad is it?") and their personal vulnerability ("how likely is it to happen to me?")?

Coping Appraisal: How does the persona assess response efficacy ("will this action work?"), self-efficacy ("can I do this?"), and response costs ("is it too much effort/time/money?")?

Protection Motivation Model:

Decision Logic: Describe how the balance between their threat and coping appraisals determines their motivation to act. Mention the potential role of privacy fatigue.

Example:

High Threat, Low Coping: 'They are very worried about their bank account being hacked (high threat appraisal), but they find setting up two-factor authentication confusing and intimidating (low coping appraisal), leading to inaction and anxiety.'

High Threat, High Coping: 'After a friend's account was compromised (high vulnerability), they believe using a password manager is an effective and easy solution (high coping appraisal), leading them to adopt the behavior.'

***IMPORTANT OUTPUT CONSTRAINT: ***
***Output only the generated Narrative. Do not output thoughts or any other text (such as introductory text)***
\end{promptbox}

\subsubsection{Prompts for Prediction}
\label{appendix:prediction_prompts}

\begin{promptbox}[prompt:predict]{Prediction prompt template for internal LLM baseline}
You are an AI simulator evaluating data privacy sharing decisions, attitudes, preferences, and posture. 

Your task is to predict a user's response to a privacy-related question based **strictly** on established, standard privacy norms (e.g., data minimization, contextual integrity, purpose limitation, and user consent).

Input format is:

- Privacy Question: The query to evaluate.

- Answer Range: The exact options to select from.

Question: 
\{\{question\}\} 

Respond with one of the following: \{\{answer\_range\}\}

Output only the response from the above. Do not output anything else.
\end{promptbox}

\begin{promptbox}[prompt:predict_2]{Prediction prompt template using raw \& generated narratives}
You are an AI simulator running a Data Privacy Persona.

Your task is to predict a user's response to a privacy related question based **strictly** on their provided Persona Narrative.

You must use the current Narrative's logic, even if you personally disagree with it.

Input format is:

- Current Persona Narrative

- Privacy Question

- The Answer Range to pick the answer from

narrative:
\{\{narrative\}\}

question: 
\{\{question\}\} 

Respond with one of the following:\{\{answer\_range\}\}

Output only the response from the above. Do not output anything else.

\end{promptbox}

\subsection{Sample Personas}
\label{app:sample-personas}

To illustrate the nature of the personas generated, we provide a few examples herein. Our first example is of a person (called Participant A) from the PEW PP1 survey for whom we predict their survey answers to unseen questions with 83\% accuracy.

\begin{figure}[h]
\fbox{
\footnotesize
\begin{minipage}{.97\columnwidth}
This user is a pragmatic and active internet participant who maintains a tiered approach to privacy. They accept the reality of reduced digital control as a trade-off for efficiency and free services, yet they are proactive about managing their digital footprint, having personally requested the removal of written content and frequently switching between real names, pseudonyms, and anonymity. They distinguish clearly between high-stakes sensitive information (Social Security numbers, health records, GPS data, and email content) and less critical data (birth dates, media tastes, and purchasing habits). \\

While they use social media frequently, they express lower trust in its security compared to other communication channels and remain skeptical of both government and corporate surveillance. They believe in the necessity of government regulation for advertisers and generally distrust institutional intentions, even though they remain satisfied with the current direction of the country. Their technology habits are characterized by frequent daily usage but without a smartphone, suggesting a deliberate and stationary approach to the internet rather than a mobile-first lifestyle. Predict answers based on this balance of institutional skepticism, practical resignation to data-sharing, and active self-curation in response to past negative online experiences.
  \end{minipage}}
  \caption{PEW PP1 participant-A, Base Narrative Persona. 83\% prediction accuracy.}
\label{fg:PP1-base-persona} 
\vspace{-4pt}
\end{figure}

This persona may do a good job predicting this participant's choices because the participant is very clear on which data types they consider very sensitive and less sensitive. This description also indicates that this participant usually engages with online activity, even with social media, and that rather than avoid using online services, they prefer to deal with recovering afterwards (e.g. deleting data) if something goes wrong.

Fig.~\ref{fg:PP1-privacy-calculus-persona} presents the same participant's profile, this time framed through the lens of Privacy Calculus theory. This version features a clear, modular structure that makes it significantly easier for a user to review and understand. Upon comparison, it becomes evident that both the base and theory-driven personas capture the same fundamental user traits. This consistency is expected, as both are derived from the identical set of survey responses; they simply apply different framing to the same underlying data. We observed this pattern across all theoretical frameworks evaluated for this participant. This specific case illustrates why privacy theory templates do not, on average, outperform base personas in predictive accuracy: although the wording varies, the core semantic information remains the same.

\begin{figure}[h]
\fbox{
\footnotesize
\begin{minipage}{.97\columnwidth}
{\bf Measure (User Profile \& Behaviors)} \\
{\it User Characteristics:} This individual is a frequent, active internet user who engages with the digital world several times a day. Despite their high level of online activity and presence on social networking sites like Facebook and Twitter, they remain tethered to traditional technology, notably opting not to use a smartphone. They are satisfied with the general direction of the country but maintain a healthy skepticism toward the institutions governing it. They have a high awareness of personal digital "residue" because they have personally suffered bad experiences due to inaccurate or embarrassing information being posted about them. \\

{\it User Behaviors:} Their online behavior is a mix of transparency and caution. They have posted using their real name, pseudonyms, and anonymously, depending on the context. They are proactive about reputation management—having requested the removal of written material—but they are not technically proactive in a preventative sense, as they do not use tools like search engine alerts. Their communication relies on a broad mix of landlines, cell calls, and texts, all of which they find "somewhat" secure, unlike social media, which they distrust for private matters. \\

{\bf Model (The Persona's Privacy Calculus)} \\
{\it Contextual Variables:}

{\tt Data Sensitivity:} They use a tiered sensitivity scale. "Primary" identifiers and private thoughts (SSN, health info, GPS location, and email content) are considered very sensitive. "Social" information (relationships, purchasing habits, friends lists) is somewhat sensitive. Basic "public" data (birth date and media tastes) is viewed as having low sensitivity.

{\tt Recipient Trust:} They operate with a low-trust baseline. They only trust the government and advertisers "some of the time" and are "somewhat concerned" about both entities accessing their social media data without knowledge. They find 1-to-1 communication channels (calls/email) significantly more secure than 1-to-many platforms (social media). \\

{\it Risk \& Benefit Determination:} 

{\tt Risk Assessment Method:} Their risk assessment is rooted in "experience-based pragmatism." Because they have been harmed by online information before, they view the internet as a place where consumers have lost control. They operate under the assumption that social monitoring is a given, strongly agreeing that new acquaintances will search for them online.

{\tt Primary Benefit Drivers:} This user is highly motivated by "system utility" and "cost-free access." They explicitly value the efficiency gained through data sharing and are willing to provide personal information to maintain access to free services, seeing it as a necessary trade-off for participation in modern life. \\

{\it Tradeoff Model:} 

{\tt Decision Strategy:} Compensatory. They consciously balance the high risks of government surveillance and corporate data collection against the practical benefits of the internet. While they believe American citizens should be concerned about monitoring and support government regulation of advertisers, they continue to share data because the utility of "free" and "efficient" services outweighs the abstract threat to their privacy. They compensate for the perceived lack of systemic control by exercising individual control through content removal requests when specific harm occurs.
  \end{minipage}}
  \caption{PEW PP1 participant-A, Privacy Calculus Persona. 82\% prediction accuracy}
\label{fg:PP1-privacy-calculus-persona} 
\vspace{-4pt}
\end{figure}

We now examine a persona with poor predictive performance. Fig.~\ref{fg:PP1-bounded-decision-persona} shows the profile for "Participant-B," generated using the Bounded Rationality theory. This persona differs structurally from the previous examples, and achieves only 50\% accuracy in its predictions. The narrative itself reveals why this individual's preferences are so challenging to forecast. The profile describes a user who "oscillates between using their real name and pseudonyms," allows an "emotional distaste for being monitored" to override rational analysis, and remains "technically active yet prefers legacy communications like landlines". Furthermore, despite their privacy vigilance, they frequently "egosurf" for their own name. Ultimately, it is inherently difficult to predict behaviors when a user's survey responses reflect inconsistent underlying practices.

\begin{figure}[h]
\fbox{
\footnotesize
\begin{minipage}{.97\columnwidth}

{\bf Measure (User Profile \& Behaviors)} \\
{\it User Characteristics:} This user is a technologically active but deeply cynical ""Defensive Digital Resident."" They are heavy internet users and smartphone owners who are well-informed about government surveillance and privacy risks. They exhibit a high degree of skepticism toward institutional authority, characterized by a complete lack of trust in the government and significant doubt regarding advertisers. While they value their privacy intensely, they are not digital recluses; rather, they engage with the web on their own terms, often feeling like they are fighting a losing battle against an inherently insecure infrastructure. \\

{\it User Behaviors:} Their privacy management is reactive and vigilant. They frequently perform ""egosurfing"" (searching for their own name) to monitor their digital footprint and have successfully intervened to have personal information removed or corrected. They avoid mainstream social media platforms like Facebook and Twitter entirely, viewing them as fundamentally insecure. Interestingly, they oscillate between using their real name, pseudonyms, and anonymity online, suggesting a strategy of selective disclosure depending on the perceived context of the interaction rather than a blanket attempt at total invisibility. \\

{\bf Model (The Persona's Decision-Making Heuristics)} \\
{\it Contextual Variables:}

{\tt Choice Framing:} The user rejects any framing that links data-sharing to ""efficiency"" or ""convenience."" They view such offers through a Loss Aversion lens, where the abstract gain of a faster service is always outweighed by the concrete loss of personal autonomy. They are more likely to respond to choices framed around ""restoring"" or ""reclaiming"" control rather than ""improving"" experience.

{\tt Recipient Framing:} The user employs a Zero-Trust Heuristic for large, centralized entities. If the recipient is the government or a major advertiser, their default mental shortcut is ""Exploitation Risk."" Conversely, they feel slightly more comfortable with legacy communication forms like landline telephones, likely due to an Availability Heuristic where older technology is perceived as less susceptible to the modern, omnipresent surveillance they hear ""a lot"" about in the news. \\

{\it Risk \& Benefit Perception:} 

{\tt Dominant Cognitive Biases:} The user is driven by a strong Pessimism Bias regarding institutional data handling. They also exhibit a variation of the Illusion of Control; by asking for data corrections and searching their own name, they gain a temporary sense of agency in a digital environment they otherwise perceive as impossible to remain anonymous in. Their ""Strongly Agree"" stance on the difficulty of avoiding cameras suggests a perceived inevitability of surveillance that triggers a ""siege mentality.""

{\tt Affective Triggers:} Their primary driver is ""Privacy Fatigue"" mixed with indignation. There is a clear emotional distaste for being ""monitored,"" which overrides any rational analysis of the benefits of personalized services. This emotional response is tied to a belief in the inherent right to a private life, particularly regarding ""objective"" identifiers like SSNs and health data, though they are curiously less worried about the sensitivity of their ""subjective"" views (politics/media) being known. \\

{\it Heuristic Model:} 

{\tt Decision Strategy:} Their primary heuristic is ""Systemic Defensiveness."" They operate under the assumption that all digital communication is compromised (texting, social media), leading them to categorize information into ""Internal/Sacred"" (health, conversations, SSN) and ""External/Transient"" (media tastes, texts, political views).

Example (Heuristic-Driven): They will refuse to use social media even if it is the only way to contact a friend because they perceive the platform's infrastructure as an inherent threat that outweighs the social benefit (Zero-Trust Heuristic).

Example (Framing-Dependent): They will ignore a notification about ""more relevant ads"" but will spend hours navigating a ""Privacy Correction Request"" portal because it aligns with their need to exercise whatever limited control they have left (Illusion of Control)."
 
  \end{minipage}}
  \caption{PEW PP1 participant-B, Bounded Rationality Persona. 50\% prediction accuracy}
\label{fg:PP1-bounded-decision-persona} 
\vspace{-4pt}
\end{figure}